\pdfoutput=1

\PassOptionsToPackage{table}{xcolor}
\PassOptionsToPackage{dvipdfm}{graphicx}
\documentclass[sigconf]{acmart}

\usepackage{stfloats}
\usepackage{tikz}
\usetikzlibrary{shapes,arrows,fit,backgrounds,positioning,matrix,calc,shapes.multipart,chains,arrows,decorations.pathreplacing,tikzmark,decorations.markings,arrows.meta,math,patterns}
\usepackage{bmpsize}
\usepackage{afterpage}
\usepackage{array}
\usepackage{tabularx}
\usepackage[utf8]{inputenc}
\usepackage{xspace}
\usepackage{csquotes}
\usepackage{./operatorsymbols}
\usepackage{numprint}
\usepackage{listings}
\usepackage{ellipsis} 
\usepackage{cleveref}
\usepackage{subcaption}
\usepackage[breakable,skins]{tcolorbox}
\usepackage{ifthen}
\usepackage{etoolbox}
\usepackage{xparse}
\usepackage{bookmark}
\npdecimalsign{.}

\AtBeginDocument{%
  }

\newcounter{revisionnumber}
\newcommand{\setrevisionnumber}[1]{\setcounter{revisionnumber}{#1}}
\newcommand{\revised}[2][]{%
  \ifx\relax#1\relax %
    {\color{blue}#2}%
  \else
    \ifnum\value{revisionnumber}<#1\relax
      {\color{blue}#2}%
    \else
      {#2}%
    \fi
  \fi
}
\setrevisionnumber{6}

\copyrightyear{2025}
\acmYear{2025}
\setcopyright{cc}
\setcctype{by-nc-sa}
\acmConference[SIGMOD-Companion '25]{Companion of the 2025 International Conference on Management of Data}{June 22--27, 2025}{Berlin, Germany}
\acmBooktitle{Companion of the 2025 International Conference on Management of Data (SIGMOD-Companion '25), June 22--27, 2025, Berlin, Germany}
\acmDOI{10.1145/3722212.3724447}
\acmISBN{979-8-4007-1564-8/2025/06}

\begin{document}

\title{Pruning in Snowflake: Working Smarter, Not Harder}

\author{Andreas Zimmerer}
\authornote{Work done while at Snowflake Inc.}
\orcid{0000-0002-4158-5805}
\affiliation{%
  \institution{University of Technology Nuremberg}
  \city{Nuremberg}
  \country{Germany}
}
\email{andreas.zimmerer@utn.de}

\author{Damien Dam}
\orcid{0009-0006-6211-4523}
\affiliation{%
  \institution{Snowflake Inc.}
  \city{Berlin}
  \country{Germany}}
\email{damien.dam@snowflake.com}

\author{Jan Kossmann}
\orcid{0000-0003-1832-7282}
\affiliation{%
  \institution{Snowflake Inc.}
  \city{Berlin}
  \country{Germany}}
\email{jan.kossmann@snowflake.com}

\author{Juliane Waack}
\orcid{0009-0002-8820-1036}
\affiliation{%
  \institution{Snowflake Inc.}
  \city{Berlin}
  \country{Germany}}
\email{juliane.waack@snowflake.com}

\author{Ismail Oukid}
\orcid{0000-0002-4253-4989}
\affiliation{%
  \institution{Snowflake Inc.}
  \city{Berlin}
  \country{Germany}}
\email{ismail.oukid@snowflake.com}

\author{Andreas Kipf}
\orcid{0000-0003-3463-0564}
\affiliation{%
  \institution{University of Technology Nuremberg}
  \city{Nuremberg}
  \country{Germany}
}
\email{andreas.kipf@utn.de}

\renewcommand{\shortauthors}{Andreas Zimmerer et al.}

\begin{abstract}

    Modern cloud-based data analytics systems must efficiently process petabytes of data residing on cloud storage.
    A key optimization technique in state-of-the-art systems like Snowflake is partition pruning---skipping chunks of data that do not contain relevant information for computing query results.
    
    While partition pruning based on query predicates is a well-established technique, we present new pruning techniques that extend the scope of partition pruning to \texttt{LIMIT}, top-k, and \texttt{JOIN} operations, significantly expanding the opportunities for pruning across diverse query types.
    We detail the implementation of each method and examine their impact on real-world workloads.
    
    Our analysis of Snowflake's production workloads reveals that real-world analytical queries exhibit much higher selectivity than commonly assumed, yielding effective partition pruning and highlighting the need for more realistic benchmarks. %
    We show that we can harness high selectivity by utilizing min/max metadata available in modern data analytics systems and data lake formats like Apache Iceberg, reducing the number of processed micro-partitions by 99.4\% across the Snowflake data platform.
  
\end{abstract}

\begin{CCSXML}
<ccs2012>
   <concept>
       <concept_id>10002951.10002952.10003190.10003192</concept_id>
       <concept_desc>Information systems~Database query processing</concept_desc>
       <concept_significance>500</concept_significance>
       </concept>
   <concept>
       <concept_id>10002951.10002952.10003190.10003192.10003210</concept_id>
       <concept_desc>Information systems~Query optimization</concept_desc>
       <concept_significance>500</concept_significance>
       </concept>
   <concept>
       <concept_id>10002951.10002952.10003190.10003191</concept_id>
       <concept_desc>Information systems~DBMS engine architectures</concept_desc>
       <concept_significance>500</concept_significance>
       </concept>
    <concept>
       <concept_id>10002951.10003152.10003517.10003176</concept_id>
       <concept_desc>Information systems~Cloud based storage</concept_desc>
       <concept_significance>500</concept_significance>
       </concept>
    <concept>
       <concept_id>10002951.10003317.10003338.10003346</concept_id>
       <concept_desc>Information systems~Top-k retrieval in databases</concept_desc>
       <concept_significance>300</concept_significance>
       </concept>
 </ccs2012>
\end{CCSXML}

\ccsdesc[500]{Information systems~Database query processing}
\ccsdesc[500]{Information systems~Query optimization}
\ccsdesc[500]{Information systems~DBMS engine architectures}
\ccsdesc[500]{Information systems~Cloud based storage}
\ccsdesc[300]{Information systems~Top-k retrieval in databases}

\keywords{data warehouses; data skipping; partition pruning; top-k queries; LIMIT pruning; join pruning; analytical query processing}

\definecolor{snowflakeblue}{HTML}{29B5E8}
\definecolor{snowflakedarkblue}{HTML}{11567F}
\definecolor{snowflakelightblue}{HTML}{70d3dc}
\definecolor{snowflakeorange}{HTML}{ff9e36}
\definecolor{snowflakelightgray}{HTML}{f2f2f2}

\definecolor{snowflakelighterblue}{HTML}{68cbef}
\definecolor{snowflakelighterdarkblue}{HTML}{5788a5}

\newcommand\paragraphnospace[1]{\noindent{\bfseries#1\space}}
\newcolumntype{C}[1]{>{\centering\arraybackslash}p{#1}}

\newcommand{\watermarked}[2]{
    \begin{tikzpicture}
        \node[opacity=1.0] at (0,0) {#1};
        \node[rotate=45, color=red, font=\Huge, opacity=0.4] at (0,0) {#2};
    \end{tikzpicture}
}
\newcommand{\draftwatermarked}[1]{\watermarked{#1}{DRAFT}}

\newcounter{n} %
\newcommand{\linebreaks}[1]{%
  \setcounter{n}{0} %
  \loop
    \ifnum\value{n}<#1 %
    \addtocounter{n}{1} %
    ~\newline %
  \repeat
}

\NewDocumentEnvironment{placeholdertextbox}{O{default}}{%
  \begin{tcolorbox}[
    breakable,
    enhanced,
    boxsep=0pt, left=0pt, right=0pt, top=0pt, bottom=0pt,
    colback=white!0!white,          %
    colframe=black!20,         %
    sharp corners,          %
    overlay={
        \draw[black!20, line width=1pt] (frame.north west) -- (frame.south east);  %
        \draw[black!20, line width=1pt] (frame.north east) -- (frame.south west);  %
    }
  ]
}{%
    \linebreaks{#1}
  \end{tcolorbox}
}

\newcommand*{\eg}{e.g.,\@\xspace}
\newcommand*{\ie}{i.e.,\@\xspace}
\newcommand*{\cf}{cf.\@\xspace}
\newcommand*{\etal}{~et~al.\@\xspace}
\newcommand*{\dash}{\textemdash\@\xspace}
\newcommand*{\etc}{~etc.\@\xspace}

\newcommand*{\s}{\,s\@\xspace}
\newcommand*{\ms}{\,ms\@\xspace}
\newcommand*{\GHz}{\,GHz\@\xspace}
\newcommand*{\B}{\,B\@\xspace}
\newcommand*{\KB}{\,KB\@\xspace}
\newcommand*{\MB}{\,MB\@\xspace}
\newcommand*{\GB}{\,GB\@\xspace}
\newcommand*{\TB}{\,TB\@\xspace}
\newcommand*{\percent}{\,\%\@\xspace}
\newcommand*{\M}{\,M\@\xspace}

\definecolor{dkgreen}{rgb}{0,0.6,0}
\definecolor{gray}{rgb}{0.5,0.5,0.5}
\definecolor{mauve}{rgb}{0.58,0,0.82}
\lstset{language=SQL,
  basicstyle={\small\ttfamily},
  belowskip=3mm,
  breakatwhitespace=true,
  breaklines=true,
  classoffset=0,
  columns=flexible,
  commentstyle=\color{dkgreen},
  framexleftmargin=0.25em,
  frameshape={}{}{}{}, %
  keywordstyle=\color{blue},
  numbers=none, %
  numberstyle=\tiny\color{gray},
  showstringspaces=false,
  stringstyle=\color{mauve},
  tabsize=3,
  xleftmargin =1em
}

\newcommand*\circled[1]{\tikz[baseline=(char.base)]{
            \node[shape=circle,draw,fill=white,inner sep=2pt] (char) {#1};}}

\pgfdeclarelayer{background}
\pgfdeclarelayer{foreground}
\pgfdeclarelayer{sankeydebug}
\pgfsetlayers{background,main,foreground,sankeydebug}

\newif\ifsankeydebug

\newenvironment{sankeydiagram}[1][]{

  \def\sankeyflow##1##2{%
    \path[sankey fill]
    let
    \p1=(##1.north east),\p2=(##1.south east),
    \n1={atan2(\y1-\y2,\x1-\x2)-90},
    \p3=(##2.north west),\p4=(##2.south west),
    \n2={atan2(\y3-\y4,\x3-\x4)+90}
    in
    (\p1) to[out=\n1,in=\n2] (\p3) --
    (\p4) to[in=\n1,out=\n2] (\p2) -- cycle;
    \draw[sankey draw]
    let
    \p1=(##1.north east),\p2=(##1.south east),
    \n1={atan2(\y1-\y2,\x1-\x2)-90},
    \p3=(##2.north west),\p4=(##2.south west),
    \n2={atan2(\y3-\y4,\x3-\x4)+90}
    in
    (\p1) to[out=\n1,in=\n2] (\p3)
    (\p4) to[in=\n1,out=\n2] (\p2);
  }

  \tikzset{
    sankey tot length/.store in=\sankeytotallen,
    sankey tot quantity/.store in=\sankeytotalqty,
    sankey min radius/.store in=\sankeyminradius,
    sankey arrow length/.store in=\sankeyarrowlen,
    sankey debug/.is if=sankeydebug,
    sankey debug=false,
    sankey flow/.style={
      to path={
        \pgfextra{
          \pgfinterruptpath
          \edef\sankeystart{\tikztostart}
          \edef\sankeytarget{\tikztotarget}
          \sankeyflow{\sankeystart}{\sankeytarget}
          \endpgfinterruptpath
        }
      },
    },
    sankey node/.style={
      inner sep=0,minimum height={sankeyqtytolen(##1)},
      minimum width=0,draw=none,line width=0pt,
    },
    sankey angle/.store in=\sankeyangle,
    sankey fill/.style={line width=0pt,fill,white},
    sankey draw/.style={draw=black,line width=.4pt},
  }

  \newcommand\sankeynode[4]{%
    \node[sankey node=##1,rotate=##2] (##3) at (##4) {};
    \ifsankeydebug
    \begin{pgfonlayer}{sankeydebug}
      \draw[red,|-|] (##3.north west) -- (##3.south west);
      \pgfmathsetmacro{\len}{sankeyqtytolen(##1)/3}
      \draw[red] (##3.west)
      -- ($(##3.west)!\len pt!90:(##3.south west)$)
      node[font=\tiny,text=black] {##3};
    \end{pgfonlayer}
    \fi
  }

  \newcommand\sankeynodestart[4]{%
    \sankeynode{##1}{##2}{##3}{##4}
    \begin{scope}[shift={(##3)},rotate=##2]
      \path[sankey fill]
      (##3.north west) -- ++(-\sankeyarrowlen,0)
      -- ([xshift=-\sankeyarrowlen/6]##3.west)
      -- ([xshift=-\sankeyarrowlen]##3.south west)
      -- (##3.south west) -- cycle;
      \path[sankey draw]
      (##3.north west) -- ++(-\sankeyarrowlen,0)
      -- ([xshift=-\sankeyarrowlen/6]##3.west)
      -- ([xshift=-\sankeyarrowlen]##3.south west)
      -- (##3.south west);
    \end{scope}
  }

  \newcommand\sankeynodeend[4]{%
    \sankeynode{##1}{##2}{##3}{##4}
    \begin{scope}[shift={(##3)},rotate=##2]
      \path[sankey fill]
      (##3.north east)
      -- ([xshift=\sankeyarrowlen]##3.east)
      -- (##3.south west) -- cycle;
      \path[sankey draw]
      (##3.north east)
      -- ([xshift=\sankeyarrowlen]##3.east)
      -- (##3.south west);
    \end{scope}
  }

  \newcommand\sankeyadvance[3][]{%
    \edef\name{##2}
    \ifstrempty{##1}{
      \def\newname{##2}
      \edef\name{##2-old}
      \path [late options={name=##2,alias=\name}];
    }{
      \def\newname{##1}
    }
    \path
    let
    \p1=(##2.north east),
    \p2=(##2.south east),
    \n1={atan2(\y1-\y2,\x1-\x2)-90},
    \p3=($(\p1)-(\p2)$),
    \n2={sankeylentoqty(veclen(\x3,\y3))},
    \p4=($(##2.east)!##3!-90:(##2.north east)$)
    in
    \pgfextra{
      \pgfmathsetmacro{\prop}{\n2}
      \pgfinterruptpath
      \sankeynode{\prop}{\n1}{\newname}{\p4}
      \path (\name) to[sankey flow] (\newname);
      \endpgfinterruptpath
    };
  }

  \newcommand\sankeyturn[3][]{%
    \edef\name{##2}
    \ifstrempty{##1}{
      \def\newname{##2}
      \edef\name{##2-old}
      \path [late options={name=##2,alias=\name}];
    }{
      \def\newname{##1}
    }
    \ifnumgreater{##3}{0}{
      \typeout{turn acw: ##3}
      \path
      let
      \p1=(##2.north east),
      \p2=(##2.south east),
      \p3=($(\p1)!-\sankeyminradius!(\p2)$),
      \n1={atan2(\y1-\y2,\x1-\x2)-90},
      \p4=($(\p1)-(\p2)$),
      \n2={sankeylentoqty(veclen(\x4,\y4))},
      \p5=(##2.east),
      \p6=($(\p3)!1!##3:(\p5)$)
      in
      \pgfextra{
        \pgfmathsetmacro{\prop}{\n2}
        \pgfinterruptpath
        \sankeynode{\prop}{\n1+##3}{\newname}{\p6}
        \path (\name) to[sankey flow] (\newname);
        \endpgfinterruptpath
      };
    }{
      \typeout{turn acw: ##3}
      \path
      let
      \p1=(##2.south east),
      \p2=(##2.north east),
      \p3=($(\p1)!-\sankeyminradius!(\p2)$),
      \n1={atan2(\y1-\y2,\x1-\x2)+90},
      \p4=($(\p1)-(\p2)$),
      \n2={sankeylentoqty(veclen(\x4,\y4))},
      \p5=(##2.east),
      \p6=($(\p3)!1!##3:(\p5)$)
      in
      \pgfextra{
        \pgfmathsetmacro{\prop}{\n2}
        \pgfinterruptpath
        \sankeynode{\prop}{\n1+##3}{\newname}{\p6}
        \path (\name) to[sankey flow] (\newname);
        \endpgfinterruptpath
      };
    }
  }

  \newcommand\sankeyfork[2]{%
    \def\name{##1}
    \def\listofforks{##2}
    \xdef\sankeytot{0}
    \path 
    let
    \p1=(\name.north east),
    \p2=(\name.south east),
    \n1={atan2(\y1-\y2,\x1-\x2)-90},
    \p4=($(\p1)-(\p2)$),
    \n2={sankeylentoqty(veclen(\x4,\y4))}
    in
    \pgfextra{
      \pgfmathsetmacro{\iprop}{\n2}
    }
    \foreach \prop/\name[count=\c] in \listofforks {
      let
      \p{start \name}=($(\p1)!\sankeytot/\iprop!(\p2)$),
      \n{nexttot}={\sankeytot+\prop},
      \p{end \name}=($(\p1)!\n{nexttot}/\iprop!(\p2)$),
      \p{mid \name}=($(\p{start \name})!.5!(\p{end \name})$)
      in
      \pgfextra{
        \xdef\sankeytot{\n{nexttot}}
        \pgfinterruptpath
        \sankeynode{\prop}{\n1}{\name}{\p{mid \name}}
        \endpgfinterruptpath
      }
    }
    \pgfextra{
      \pgfmathsetmacro{\diff}{abs(\iprop-\sankeytot)}
      \pgfmathtruncatemacro{\finish}{\diff<0.01?1:0}
      \ifnumequal{\finish}{1}{}{
        \message{*** Warning: bad sankey fork (maybe)...}
        \message{\iprop-\sankeytot}
      }
    };
  }

  \tikzset{
    declare function={
      sankeyqtytolen(\qty)=\qty/\sankeytotalqty*\sankeytotallen;
      sankeylentoqty(\len)=\len/\sankeytotallen*\sankeytotalqty;
    },
    sankey tot length=100pt,
    sankey tot quantity=100,
    sankey min radius=30pt,%
    sankey arrow length=10pt,%
    #1}
}{
}

\maketitle

\section{Introduction}
\label{s:introduction}

The fastest way of processing data is to not process it at all.
Excluding chunks of data, \ie \emph{pruning} them, is indispensable for query performance.
In cloud-based data analytics systems, pruning becomes even more important:
such systems decouple compute and storage. Hence, pruning serves not only to avoid processing and loading data from disk but is also essential in avoiding costly network I/O.
Given that scan and filter operations typically account for more than 50\% of processing time~\cite{vanRenen2023}, effective data pruning becomes a key driver for query performance in cloud-based data analytics systems.
For instance, Google's PowerDrill attributes its performance to skipping 92.41\% of data on average~\cite{Hall2012}. Snowflake, in turn, has achieved a staggering 99.4\% of data being skipped across all customer workloads\footnote{Measured from Nov 18\textsuperscript{th} to Nov 24\textsuperscript{th} 2024.}. Though not directly comparable\footnote{PowerDrill performs an initial pruning pass on a table's clustering-key, which is not reflected in this number.}, both figures underscore the importance of data skipping.

\begin{figure}[t]
    \vspace{1.5em}
    \centering
    \includegraphics[width=\linewidth]{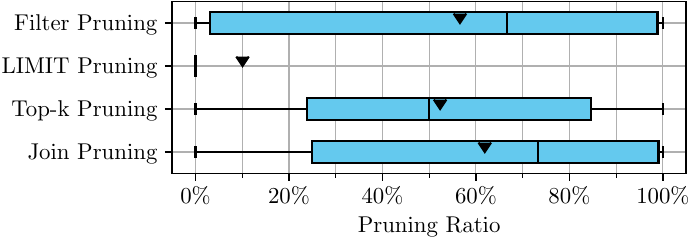}
    \caption{Pruning ratios of different pruning techniques for eligible queries in Snowflake. $\blacktriangledown$ indicates mean values.
    Based on representative samples across all customers from Nov 5\textsuperscript{th} to Nov 7\textsuperscript{th} 2024.
    }
    \label{fig:intro-fig}
\end{figure}

Data in cloud-based OLAP systems is typically broken down into implicit, horizontal \emph{micro-partitions} that can be accessed independently by the query engine.
Using partition-level metadata, an engine can skip over micro-partitions that do not contain any qualifying tuples.
Besides simple min-max values per column (zone maps~\cite{Graefe2009} or SMAs~\cite{Moerkotte1998}), there is a vast array of techniques that improve data skipping.
These include secondary index structures~\cite{kipf2020cuckoo}, approximate set-membership data structures (\eg Bloom-filters~\cite{bloom1970space, lang2019performance}, Cuckoo-filters~\cite{fan2014cuckoo}, or Xor-filters~\cite{graf2020xor, dietzfelbinger2008succinct}), and optimizations for joins through sideways information passing~\cite{mackert1986r, mullin1990optimal}.

In the context of cloud-based systems with decoupled compute and storage architectures, the high latency and costs associated with network I/O amplify the importance of accurate, aggressive, and efficient data pruning.
Skipping irrelevant data early in the query life-cycle prevents not just unnecessary data loads from remote storage when scanning data, but also reduces communication overhead between nodes when exchanging information on what data needs to be scanned.
Besides its highly optimized proprietary format, Snowflake also supports the Apache Iceberg~\cite{icebergformat} open table format with the Apache Parquet file format~\cite{parquetformat}.
All techniques presented in this paper are also applicable to Iceberg tables in Snowflake which will be further discussed in \Cref{s:pruning-for-iceberg}.

Despite significant research in improving pruning through auxiliary metadata and indexing techniques, these approaches often operate under idealized assumptions.
Many fail to account for complex, real-world scenarios. For instance, Ives and Taylor evaluated their \emph{sideways information passing} technique~\cite{ives2008sideways} using a modified TPC-H benchmark.
However, as we demonstrate in \Cref{s:tpch}, TPC-H exhibits significantly different pruning behavior compared to real-world workloads.

The objective of this paper is to provide insights into the pruning process and its efficacy on real-world workloads in large cloud-based data systems, such as Snowflake.
We examine how the decoupled compute and storage architecture influences pruning design, evaluate various pruning techniques in real-world settings, and propose novel enhancements to improve query performance.
\Cref{fig:intro-fig} illustrates the pruning ratios of various specialized techniques in Snowflake, demonstrating their substantial impact. Although few queries benefit from \texttt{LIMIT} pruning, its impact on those is significant, as shown by the high mean relative to the low median.

It is worth noting that, regardless of the implemented pruning techniques, the number of data partitions that can be skipped primarily depends on how data is distributed among micro-partitions, \eg via sorting or clustering, and on the concrete workload.
Data layout optimizations and adaptive partitioning strategies based on workload characteristics for better pruning (\eg\cite{yang2020qd, wu2011partitioning, ceri1982horizontal, agrawal2004integrating, sun2016skipping, athanassoulis2019optimal, ding2021instance}) are topics on their own and beyond the scope of this paper.

\medskip
\noindent Our contributions are as follows:
\begin{itemize}
    \item We evaluate and quantify the impact of four different pruning techniques: filter pruning, \texttt{LIMIT} pruning, top-k pruning, and \texttt{JOIN} pruning.
    \item We propose a novel technique for pruning LIMIT queries, both with and without predicates.
    \item We demonstrate how pruning techniques traditionally used in search engines for top-k queries can be adapted and integrated into SQL engines, yielding significant performance improvements. Additionally, we propose optimizations to further enhance the effectiveness of these techniques.
    \item Unlike previous works that focus predominantly on single pruning techniques in isolation, we illustrate their combined effectiveness through a guiding example and showcase how they complement each other.
\end{itemize}
\medskip

\noindent The paper is structured as follows:
We first provide an architectural overview of the Snowflake Data Platform in \Cref{s:snowflake-architecture}, outlining how pruning integrates into and is influenced by the platform’s overarching design, as detailed in \Cref{s:pruning-at-snowflake}.
Next, we examine several pruning techniques implemented in Snowflake, assessing their effectiveness and impact.
Specifically, \Cref{s:filter-pruning} demonstrates how filter pruning effectively reduces the number of micro-partitions processed for queries with predicates.
In \Cref{s:limit-pruning}, we explore a specialized pruning technique tailored to \texttt{LIMIT} queries, leveraging extensions to filter pruning to extract meaningful table structure insights. 
For \emph{top-k} queries, \Cref{s:topk-pruning} introduces a pruning strategy inspired by concepts from the Information Retrieval~(IR) domain, achieving substantial performance improvements.
\Cref{s:join-pruning} focuses on partition pruning for \texttt{JOIN} queries.
Finally, in \Cref{s:discussion}, we discuss pruning in the context of data lake environments and critically examine the limitations of synthetic benchmarks such as TPC-H in accurately capturing the impact of partition pruning.

\section{Architecture of Snowflake}
\label{s:snowflake-architecture}

Snowflake is a cloud-based data platform, offering, among other services, scalable analytics across multiple cloud providers as a service for enterprise customers.
Its foundational design, as detailed in the original Snowflake paper~\cite{Dageville2016}, introduced the \emph{multi-cluster, shared data architecture}: storage is decoupled from a pool of compute nodes organized in a shared-nothing setup.
Together with a dedicated metadata store, this architecture has, more than a decade later, stood the test of time and has influenced the structure of other cloud-based data systems.
At Snowflake, this design shapes the development of essential components, such as pruning.

\begin{figure}[t]
    \centering
    \tikzset{
        filegroup2/.pic={
            \node[draw, rounded corners, minimum width=0.7cm, minimum height=0.6cm, fill=lightgray] (file1) at (0,0) {};
            \node[draw, rounded corners, minimum width=0.7cm, minimum height=0.6cm, fill=lightgray] (file2) at (0.15,-0.15) {};
            \node[draw, rounded corners, minimum width=0.7cm, minimum height=0.6cm, fill=lightgray] (file3) at (0.3,-0.3) {};
        }
    }
    \begin{tikzpicture}[baseline, x=1cm, y=1cm]
        \node[draw, dashed, rounded corners, minimum width=7.35cm, minimum height=1.85cm](CloudServicesLayer) {};
            \node[right=.7em of CloudServicesLayer.east, rotate=-90, anchor=base, align=center] {Services};
         \node[draw, minimum height=0.5cm, anchor=north west, align=center, font=\scriptsize, minimum width=4.85cm] (auth) at ($(CloudServicesLayer.north west)+(5pt,-5pt)$) {Authentication and Access Control};
         \node[draw, minimum height=1.51cm, anchor=north east, align=center, font=\scriptsize] (metadata) at ($(CloudServicesLayer.north east)+(-5pt, -5pt)$) {Metadata\\Storage};
         \node[draw, minimum height=0.5cm, minimum width=1.5cm, anchor=north east, align=center, font=\scriptsize] (optimizer) at ($(auth.south east)+(0pt, -3pt)$) {Optimizer};
         \node[draw, minimum height=0.3cm, minimum width=1.5cm, anchor=north west, align=center, font=\scriptsize, inner sep=1pt] (coordinator) at ($(optimizer.south west)+(0pt, -2pt)$) {Coordinator};
         \node[draw, minimum height=0.88cm, anchor=north west, align=center, font=\scriptsize] (infra) at ($(auth.south west)+(0pt, -3pt)$) {Infrastructure\\Manager};
         \node[draw, minimum height=0.88cm, right=3pt of infra.north east, anchor=north west, align=center, font=\scriptsize] (trx) {Transaction\\Manager};
         \node[draw, minimum height=0.88cm, right=3pt of trx.north east, anchor=north west, align=center, font=\scriptsize] {...};

        \node[draw, dashed, rounded corners, minimum width=7.35cm, minimum height=1.4cm, below=0.3cm of CloudServicesLayer.south](VirtualWarehousesLayer) {};
            \node[right=.7em of VirtualWarehousesLayer.east, rotate=-90, anchor=base, align=center] {Compute};

        \node[draw, minimum width=1.4cm, minimum height=1.1cm] (warehouse1) at ($(VirtualWarehousesLayer.north west)+(4.6,-0.7)$) {};
            \node[below=0em of warehouse1.north, anchor=north, align=center, font=\scriptsize] {Compute Pool};
            \node[draw, minimum width=6pt, minimum height=6pt, inner sep=0pt, anchor=north west] (worker1) at ($(warehouse1.north west)+(4pt,-12pt)$) {};
            \node[draw, minimum width=6pt, minimum height=6pt, inner sep=0pt, anchor=north west] (worker2) at ($(warehouse1.north west)+(13pt,-12pt)$) {};
            \node[draw, minimum width=6pt, minimum height=6pt, inner sep=0pt, anchor=north west] (worker3) at ($(warehouse1.north west)+(22pt,-12pt)$) {};
            \node[draw, minimum width=6pt, minimum height=6pt, inner sep=0pt, anchor=north west] (worker4) at ($(warehouse1.north west)+(31pt,-12pt)$) {};
        \node[draw, minimum width=1.4cm, minimum height=1.1cm] (warehouse2) at ($(VirtualWarehousesLayer.north west)+(6.4,-0.7)$) {};
            \node[below=0em of warehouse2.north, anchor=north, align=center, font=\scriptsize] {Compute Pool};
            \node[draw, minimum width=6pt, minimum height=6pt, inner sep=0pt, anchor=north west] (worker5) at ($(warehouse2.north west)+(4pt,-12pt)$) {};
            \node[draw, minimum width=6pt, minimum height=6pt, inner sep=0pt, anchor=north west] (worker6) at ($(warehouse2.north west)+(13pt,-12pt)$) {};
        \node[draw, minimum width=1.4cm, minimum height=1.1cm] (warehouse3) at ($(VirtualWarehousesLayer.north west)+(2.8,-0.7)$) {};
            \node[below=0em of warehouse3.north, anchor=north, align=center, font=\scriptsize] {Compute Pool};
            \node[draw, minimum width=6pt, minimum height=6pt, inner sep=0pt, anchor=north west] (worker7) at ($(warehouse3.north west)+(4pt,-12pt)$) {};
            \node[draw, minimum width=6pt, minimum height=6pt, inner sep=0pt, anchor=north west] (worker8) at ($(warehouse3.north west)+(13pt,-12pt)$) {};
            \node[draw, minimum width=6pt, minimum height=6pt, inner sep=0pt, anchor=north west] (worker9) at ($(warehouse3.north west)+(22pt,-12pt)$) {};
            \node[draw, minimum width=6pt, minimum height=6pt, inner sep=0pt, anchor=north west] (worker10) at ($(warehouse3.north west)+(31pt,-12pt)$) {};
            \node[draw, minimum width=6pt, minimum height=6pt, inner sep=0pt, anchor=north west] (worker11) at ($(warehouse3.north west)+(4pt,-21pt)$) {};
            \node[draw, minimum width=6pt, minimum height=6pt, inner sep=0pt, anchor=north west] (worker12) at ($(warehouse3.north west)+(13pt,-21pt)$) {};
            \node[draw, minimum width=6pt, minimum height=6pt, inner sep=0pt, anchor=north west] (worker13) at ($(warehouse3.north west)+(22pt,-21pt)$) {};
            \node[draw, minimum width=6pt, minimum height=6pt, inner sep=0pt, anchor=north west] (worker14) at ($(warehouse3.north west)+(31pt,-21pt)$) {};
        \node[draw, minimum width=1.4cm, minimum height=1.1cm] (warehouse4) at ($(VirtualWarehousesLayer.north west)+(1.0,-0.7)$) {};
            \node[below=0em of warehouse4.north, anchor=north, align=center, font=\scriptsize] {Compute Pool};
            \node[draw, minimum width=6pt, minimum height=6pt, inner sep=0pt, anchor=north west] (worker15) at ($(warehouse4.north west)+(4pt,-12pt)$) {};
            \node[draw, minimum width=6pt, minimum height=6pt, inner sep=0pt, anchor=north west] (worker16) at ($(warehouse4.north west)+(13pt,-12pt)$) {};

        \draw[-latex] ($(optimizer.east)+(0pt,2pt)$) to[out=20,in=160,distance=10pt] ($(metadata.west)-(0pt,2pt)$);
        \draw[-latex] ($(metadata.west)-(0pt,6pt)$) to[out=-160,in=-20,distance=10pt] ($(optimizer.east)-(0pt,2pt)$);

        \draw[-latex] ($(coordinator.south)-(2pt,0pt)$) to[out=-110,in=110,distance=7pt] ($(warehouse1.north)-(2pt,0pt)$);
        \draw[latex-] ($(coordinator.south)+(2pt,0pt)$) to[out=-70,in=70,distance=7pt] ($(warehouse1.north)+(2pt,0pt)$);

        \node[draw, dashed, rounded corners, minimum width=7.35cm, minimum height=1.1cm, below=0.3cm of VirtualWarehousesLayer.south](DataStorageLayer) {};
            \node[right=.7em of DataStorageLayer.east, rotate=-90, anchor=base, align=center] {Storage};

        \node[draw=none, fill=none, anchor=north, inner sep=0pt, outer sep=0pt] (fil1) at ($(DataStorageLayer.north west)+(0.5cm,-0.2cm)$) {\includegraphics[width=0.7cm]{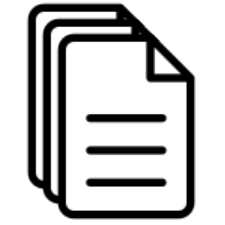}};
        \node[draw=none, fill=none, anchor=north, inner sep=0pt, outer sep=0pt] (fil2) at ($(DataStorageLayer.north west)+(1.3cm,-0.2cm)$) {\includegraphics[width=0.7cm]{figures/icons/file_stack.pdf}};
        \node[draw=none, fill=none, anchor=north, inner sep=0pt, outer sep=0pt] (fil3) at ($(DataStorageLayer.north west)+(2.1cm,-0.2cm)$) {\includegraphics[width=0.7cm]{figures/icons/file_stack.pdf}};
        \node[draw=none, fill=none, anchor=north, inner sep=0pt, outer sep=0pt] (fil4) at ($(DataStorageLayer.north west)+(2.9cm,-0.2cm)$) {\includegraphics[width=0.7cm]{figures/icons/file_stack.pdf}};
        \node[draw=none, fill=none, anchor=north, inner sep=0pt, outer sep=0pt] (fil5) at ($(DataStorageLayer.north west)+(3.7cm,-0.2cm)$) {\includegraphics[width=0.7cm]{figures/icons/file_stack.pdf}};
        \node[draw=none, fill=none, anchor=north, inner sep=0pt, outer sep=0pt] (fil6) at ($(DataStorageLayer.north west)+(4.5cm,-0.2cm)$) {\includegraphics[width=0.7cm]{figures/icons/file_stack.pdf}};
        \node[draw=none, fill=none, anchor=north, inner sep=0pt, outer sep=0pt] (fil7) at ($(DataStorageLayer.north west)+(5.3cm,-0.2cm)$) {\includegraphics[width=0.7cm]{figures/icons/file_stack.pdf}};
        \node[draw=none, fill=none, anchor=north, inner sep=0pt, outer sep=0pt] (fil8) at ($(DataStorageLayer.north west)+(6.1cm,-0.2cm)$) {\includegraphics[width=0.7cm]{figures/icons/file_stack.pdf}};
        \node[draw=none, fill=none, anchor=north, inner sep=0pt, outer sep=0pt] (fil9) at ($(DataStorageLayer.north west)+(6.9cm,-0.2cm)$) {\includegraphics[width=0.7cm]{figures/icons/file_stack.pdf}};

        \draw[-latex] (fil1.north) to[out=90,in=-90, out min distance=0.7cm,  looseness=0.3, in min distance=1.4cm] (worker1.south);
        \draw[-latex] (fil4.north) to[out=90,in=-90, out min distance=0.7cm,  looseness=0.3, in min distance=1.4cm] (worker1.south);
        \draw[-latex] (fil5.north) to[out=90,in=-90, out min distance=0.7cm,  looseness=0.3, in min distance=1.4cm] (worker2.south);
        \draw[-latex] (fil7.north) to[out=90,in=-90, out min distance=0.7cm,  looseness=0.3, in min distance=1.4cm] (worker3.south);
        \draw[-latex] (fil8.north) to[out=90,in=-90, out min distance=0.7cm,  looseness=0.3, in min distance=1.4cm] (worker4.south);

    \end{tikzpicture}
    \caption{Architecture of the Snowflake Data Platform.}
    \label{fig:snowflake-architecture}
\end{figure}

In the following paragraphs, we will briefly give an overview of the main components of the Snowflake Data Platform, which are also illustrated in \Cref{fig:snowflake-architecture}.

\medskip
\paragraphnospace{Data Storage.}
Snowflake leverages cloud object storage, \eg AWS S3~\cite{awss3}, Azure Blob Storage~\cite{azsureblob}, or GCP Cloud Storage~\cite{gcpstorage}, as its disaggregated storage layer.
Tables are implicitly horizontally partitioned at row boundaries at different granularities depending on the table format.
Regular Snowflake tables are split into \emph{micro-partitions}, typically containing between 50 and 500\MB of uncompressed data. Tables in the Apache Iceberg table format backed by Apache Parquet files follow a similar pattern, but the partition sizes depend on the writer of the file.
Both adhere to a PAX-style storage layout~\cite{ailamaki2001weaving}.
Additional information about Iceberg tables in Snowflake is discussed in \Cref{s:pruning-for-iceberg}.

\medskip
\paragraphnospace{Virtual Warehouses.}
Virtual warehouses provide the compute resources for query processing and are ephemeral, user-specific resources.
Each virtual warehouse comprises a fleet of up to several hundred compute nodes organized in a shared-nothing architecture, facilitating highly parallel query execution.
When a query is scheduled for execution, the warehouse receives an optimized query plan that contains \emph{scan sets}, \ie a serialized list of micro-partition identifiers to be processed as part of the query.
Note that a scan set might have already been subject to partition pruning at query compilation time, and a compute node in the virtual warehouse might further prune the scan set before loading data as some pruning techniques can only be executed at query runtime.

\medskip
\paragraphnospace{Cloud Services.}
Snowflake's control plane operates as a highly multi-tenant service layer.
It processes incoming queries, performs compilation and optimization steps and schedules queries for execution on virtual warehouses.
Part of Snowflake’s cloud services is a dedicated metadata service: a scalable, transactional key-value store that manages metadata for partitions stored in the data storage layer.
This design enables fast, independent access to metadata without loading the actual data, which is essential for fast and effective partition pruning during query optimization.
Additionally, the control plane handles access control, transaction management, encryption keys, administration of virtual warehouses, and so forth.

\subsection{Pruning at Snowflake}
\label{s:pruning-at-snowflake}

Pruning, also known as data skipping~\cite{Sun2014}, is a common technique in data analytics systems to minimize the volume of data being processed.
In Snowflake, pruning is primarily performed at the level of \emph{micro-partitions}; for tables in Apache Parquet format, it occurs at the file, row-group, and page levels.
However, for simplicity we use the term \emph{(micro-)partition} interchangeably throughout this paper.
Snowflake performs min/max pruning based on lightweight metadata maintained for each micro-partition, similar to zone-maps~\cite{Graefe2009} or small-materialized-aggregates (SMAs)~\cite{Moerkotte1998}.
By comparing this metadata against the query’s predicates, the database engine can efficiently identify micro-partitions that do not contain relevant data, allowing them to be skipped entirely.

Consider, for instance, two micro-partitions $f_1$ and $f_2$ containing the value ranges \texttt{0..9} and \texttt{10..19}, respectively.
If a query contains the predicate \texttt{WHERE x >= 15}, the query engine can skip $f_1$ because its maximum value, \texttt{9}, is less than \texttt{15} and therefore does not contain any matching data.
Conversely, $f_2$ must be processed because the predicate's value \texttt{15} falls within its range of \texttt{10..19}.

Pruning is designed to guarantee no false negatives, meaning it guarantees correctness by ensuring all relevant data is included in the query result.
It might, however, still produce false positives because sometimes a micro-partition deemed relevant does actually not contain matching data.
At Snowflake, partition pruning is performed in various places depending on the query type, but it can generally be categorized into two phases: \emph{compile-time} pruning and \emph{runtime} pruning.

\medskip
\paragraphnospace{Compile-time pruning} happens during query compilation and is therefore driven by the cloud services layer.
Performing pruning at such an early stage in the query life cycle has the advantage that the pruning result can be leveraged during query compilation for further optimizations, potentially resulting in a better query plan.
These optimizations can range from more accurate cardinality estimates---and therefore better join ordering---to the elimination of entire sub-trees in the query plan.
To allow pruning during compilation, fast access to micro-partition metadata is essential.

\medskip
\paragraphnospace{Runtime pruning} happens during query execution on the execution layer in virtual warehouses.
This implies that pruning algorithms need to be designed for high degrees of parallelism and with minimal inter-node communication.
Despite these constraints, runtime pruning offers two main advantages:
First, pruning is performed in parallel on multiple machines, which can make it beneficial to dynamically push compile-time pruning to a virtual warehouse if pruning itself is time-consuming. %
Second, while compile-time pruning is limited to \emph{static} pruning, pruning at runtime can incorporate information that becomes available only during query execution, allowing data-dependent pruning techniques like top-k pruning (see \Cref{s:topk-pruning}) or join pruning (see \Cref{s:join-pruning}).
Such pruning techniques require a flexible execution engine capable of passing information both horizontally and vertically.

\medskip
\paragraphnospace{Summary.}
\noindent In conclusion, effective pruning has multiple benefits:
\begin{enumerate}
    \item Faster table scans with significantly reduced data needing to be loaded over network or from disk.
    \item Enhanced cardinality estimation, leading to better join ordering and overall improved query plan quality.
    \item More accurate work estimation, enhancing the precision of query scheduling.
    \item Due to Snowflake's architecture, effective pruning results in smaller \emph{scan sets}, reducing the (de)serialization work and resulting in less data transported over the network, especially for large tables.
\end{enumerate}

\medskip
In the following sections, we will examine four types of pruning techniques in the Snowflake Data Platform and analyze their impact on customer workloads.

\section{Filter Pruning}
\label{s:filter-pruning}

Min-max pruning for filter predicates represents the traditional approach to pruning in data analytics systems~\cite{Sun2014}.
The concept is simple: Using the query's predicates, the query engine attempts to deduce whether a micro-partition might contain relevant data based on the partition's metadata. 
If it can conclusively confirm that a partition does not contain any rows that satisfy the predicates, the partition's identifier is removed from the \emph{scan set}.
Despite its widespread adoption and a general agreement of its effectiveness, we found no studies that measure the impact of this technique on real-world workloads.
We present such an analysis in \Cref{s:filter-pruning-impact}.

Snowflake performs filter pruning both during compilation and query runtime, carefully optimizing the balance between these phases to achieve best overall query execution times.
Partition pruning at compilation time offers the advantage that the pruning result can be utilized for further plan optimizations, such as improved cardinality estimates for join ordering, query sub-tree elimination, constant folding, and other cost-based decisions, potentially leading to improved query plans.
Filter pruning at compile time is designed as a dynamic multi-step process: as new filters are identified, the pruning results are incrementally refined.

As an illustrative example, we use the scenario of the International Union for Conservation of Nature (IUCN)\footnote{\url{https://iucn.org/}} identifying suitable locations for an animal observation post on a mountain ridge.
Beyond this data analysis case, similar examples can be envisioned in areas such as cybersecurity or process monitoring \etc

Initially, a data analyst may aim to list all potential locations, specifically those situated along an existing trail on a mountain ridge above a specified altitude. Since the altitude could be provided in either feet or meters, it must first be converted to a consistent unit.
A possible query for this scenario could look like the following:
\begin{lstlisting}[label=l:filter-example,language=SQL]
SELECT * FROM trails
WHERE IF(unit='feet', altit * 0.3048, altit) > 1500 
  AND name LIKE 'Marked-%-Ridge';
\end{lstlisting}
\noindent Based on this example, we examine two key aspects: pruning complex expressions and balancing compile-time and runtime pruning.

\subsection{Pruning Complex Expressions}\label{s:filter-pruning-complex-expressions}

Pruning micro-partitions based on simple base-value predicates is not enough for most queries.
In reality, many predicates make use of unary---or even $n$-ary functions involving multiple columns---and perform non-trivial computations with their parameters as we can also see in our example query.
To further illustrate the scenario, we consider the following available metadata:

\begin{table}[H]
    \begin{tabular}[t]{l|C{2cm}|C{2cm}}
        \hline
        \textbf{Column} & \textbf{Min}   & \textbf{Max} \\\hline
        \texttt{unit}   & \enquote{feet} & \enquote{meters} \\
        \texttt{altit}  &     934        &    7674          \\
        \texttt{name}   & \enquote{Basecamp-...} & \enquote{Unmarked-...} \\
        \hline
    \end{tabular}
\end{table}

\noindent How can the predicate in our example be used to efficiently prune micro-partitions when only basic min/max metadata is available?
To address this, consider each component of the predicate individually.

\medskip
\paragraphnospace{Deriving Min/Max Ranges.}
For \lstinline{(altit * 0.3048)}, the original range of the \texttt{altit} column is scaled, resulting in a transformed range of around \texttt{(min=284.68, max=2339.04)}.
For the \lstinline{IF(...)}, where it is not always possible to determine which values in \lstinline{unit} are equal to \lstinline{'feet'}, a conservative approach must be adopted.
Here, this means that the resulting min/max range is extended to encompass the min/max ranges of both sub-expressions, yielding \texttt{(min=284.68, max=7674)}. If there are micro-partitions where the min/max metadata indicates that either none or all values of \lstinline{unit} are equal to \lstinline{'feet'}, the ranges can be adjusted accordingly.
The comparison \lstinline{>1500} partially overlaps with the range \texttt{(min=284.68, max=7674)}, meaning the first boolean expression evaluates to \lstinline{true}. This indicates the micro-partition \emph{might} contain rows satisfying the predicate.
For effective pruning, every function must provide a mechanism to derive transformed min/max ranges from its input.

\medskip
\paragraphnospace{Imprecise Filter Rewrites.}
For the second filter \lstinline|(name LIKE ('Marked-%-Ridge'))|, we employ a technique we refer to as \emph{imprecise filter rewrite}.
While predicates can only be rewritten to semantically equivalent expressions to ensure correctness for query evaluation, pruning allows for a different approach: predicates can be \emph{widened} to facilitate more coarse-grained pruning.
This enables predicates that are otherwise not directly applicable for pruning to be transformed into less precise forms that support pruning, as demonstrated in this case.
For \lstinline|'Marked-%-Ridge'|, the predicate can be widened to \lstinline{STARTSWITH('Marked-')}, effectively relaxing the constraint that the string has to end with \lstinline|'-Ridge'|.
Pruning for \lstinline|STARTSWITH| is then performed by checking whether the string falls within the min/max range of the column.

\noindent The resulting expression used for pruning micro-partitions is:
{\small
\begin{align*}
    &\texttt{name}_{min} <= \text{\lstinline{'Marked-'}} <= \texttt{name}_{max} \land (\\
    &\quad((\texttt{altit}_{min} = \text{\lstinline{'feet'}} \land \texttt{altit}_{max} = \text{\lstinline{'feet'}}) \rightarrow \\
    &\qquad \texttt{altit}_{max} * 0.3048 > 1500) \lor \\
    &\quad((\texttt{altit}_{max} < \text{\lstinline{'feet'}} \lor \texttt{altit}_{min} > \text{\lstinline{'feet'}}) \rightarrow\\
    &\qquad \texttt{altit}_{max} > 1500) \lor \\
    &\quad((\texttt{altit}_{max} \geq \text{\lstinline{'feet'}} \land \texttt{altit}_{min} \leq \text{\lstinline{'feet'}}) \rightarrow\\
    &\qquad \max(\texttt{altit}_{max} * 0.3048, \texttt{altit}_{max}) > 1500) )
\end{align*}
}%
Evaluating this expression against the provided metadata in the example indicates that the micro-partition should not be pruned.

\subsection{Balancing Compile-Time and Runtime Filter Pruning}\label{s:balanceing-compiletime-runtime-filter-pruning}

As previously noted, compile-time pruning introduces additional overhead to the compilation process, which can become prohibitively expensive for queries on extremely large tables with complex and deeply nested predicates.
To maintain fast query execution times, Snowflake employs two key strategies:
(i)~reordering the evaluation sequence of filters during partition pruning to prioritize fast and effective filters and
(ii)~halting pruning for filters that prove ineffective or slow during compile time, with the option of deferring their execution to the highly parallelized virtual warehouses and hence pushing their evaluation to query execution time. \\
Both approaches rely on monitoring the partition pruning ratio for each filter as well as the average time required to evaluate the filtering predicate on a micro-partition.

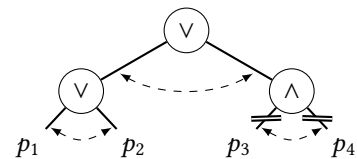
\begin{figure}[htb]
    \centering
    \tikzset{
        pruner/.style={text centered, inner sep=3, align=center,font=\large, minimum height=1.5em},
        strike through right/.style={
            postaction=decorate,
            decoration={
                markings,
                mark=at position 0.5 with {
                    \draw[-,double] (-4pt,-4pt) -- (4pt,4pt);
                }
            }
        },
        strike through left/.style={
            postaction=decorate,
            decoration={
                markings,
                mark=at position 0.5 with {
                    \draw[-,double] (-4pt,4pt) -- (4pt,-4pt);
                }
            }
        }
    }
    \begin{tikzpicture}[baseline={(current bounding box.north)}]
        \def\xdir{14}
        \def\ydir{2}

        \node (l0or) [circle, pruner, draw] {$\lor$};
        \node (l1or) [circle, pruner, draw, below=\ydir*1mm of l0or, xshift=\xdir*-1mm] {$\lor$};
        \node (l1and) [circle, pruner, draw, below=\ydir*1mm of l0or, xshift=\xdir*1mm] {$\land$};
        \node (p1) [pruner, below=\ydir*1mm of l1or, xshift=\xdir*-.5mm] {$p_1$};
        \node (p2) [pruner, below=\ydir*1mm of l1or, xshift=\xdir*.5mm] {$p_2$};
        \node (p3) [pruner, below=\ydir*1mm of l1and, xshift=\xdir*-.5mm] {$p_3$};
        \node (p4) [pruner, below=\ydir*1mm of l1and, xshift=\xdir*.5mm] {$p_4$};

        \draw[thick] (l0or) -- (l1or);
        \draw[thick] (l0or) -- (l1and);
        \draw[thick] (l1or) -- (p1);
        \draw[thick] (l1or) -- (p2);
        \draw[thick,strike through left] (l1and) -- (p3);
        \draw[thick,strike through right] (l1and) -- (p4);

        \draw[latex-latex,dashed] ($(l1or.north east) + (.3cm, .02cm)$) to[out=-30,in=210,distance=.6cm] ($(l1and.north west) + (-.3cm, .02cm)$);
        \draw[latex-latex,dashed] ($(p1.north east) + (.02cm, .02cm)$) to[out=-30,in=210,distance=.4cm] ($(p2.north west) + (-.02cm, .02cm)$);
        \draw[latex-latex,dashed] ($(p3.north east) + (.02cm, .02cm)$) to[out=-30,in=210,distance=.4cm] ($(p4.north west) + (-.02cm, .02cm)$);
    \end{tikzpicture}
    \caption{Exemplary pruning-tree with predicates $p_i$ for the expression $(p_1 \lor p_2) \lor (p_3 \land p_4)$ showing re-ordering and cutoff options. Cutoff can only be performed below an $\land$.}
    \label{fig:pruner-tree}
\end{figure}

\medskip
\paragraphnospace{Filter Reordering.}
For (i), when pruning involves multiple filter predicates, these predicates are connected by boolean operators, forming a tree-like structure.
An example for this is depicted in \Cref{fig:pruner-tree}.
The filter predicates serve as the leaves, while the boolean operators ($\land$ or $\lor$) act as the inner nodes.
The order in which the direct children of $\land$ and $\lor$ nodes are evaluated can be rearranged freely (dashed arrows in \Cref{fig:pruner-tree}).
This flexibility allows prioritizing fast and highly selective filters for $\land$ expressions, as they quickly reduce the number of micro-partitions.
Conversely, for $\lor$ expressions, it is better to prioritize fast filters with low selectivity.
An initial pruning order can be established heuristically. As pruning progresses, Snowflake tracks the pruning ratio and evaluation time for each node in the pruning tree, making local adjustments to improve efficiency. Tracking these metrics incurs little overhead and gradually performing local decisions ensures the complexity of reordering does not explode. Over time, this adaptive, localized approach may lead to a more optimal global execution order.

\medskip
\paragraphnospace{Filter Pruning Cutoff.}
If reordering filters during pruning does still not meet the desired performance requirements, strategy (ii), \ie halting pruning for slow or ineffective filters, is applied.
We call this \enquote{filter pruning cutoff} and apply this to filters that are either slow, or ineffective, or both.
Similar to reordering, Snowflake monitors execution speed and pruning ratio for each node in the pruning-tree and repeatedly check for nodes violating the performance requirements. A conservative, but simple, metric for locally checking if a pruning node is slow or ineffective is to model and compare two scenarios:
We can either continue using the filter for pruning and extrapolate its pruning ratio to the remaining \emph{scan set}, considering its computational overhead.
Or we can estimate the impact on overall query performance if we would stop pruning with that filter and let the parallel execution engine process all remaining micro-partitions.
The scenario yielding the better prediction regarding overall query execution time should be chosen.
However, it is important to note that this simple metric does not account for specific query characteristics; for instance, removing a pruner might prevent other optimizations from being applied, \eg sub-tree elimination or constant-folding.

Additionally, care must be taken when removing pruning filters from the predicate tree to avoid incorrect results.
When deciding to stop pruning for a filter, a conservative approach should be used, i.e., we should assume that every micro-partition passes that filter from the point on where we stop evaluating it.
If a filter is part of an $\land$-expression, it can be safely removed since its role is to further narrow the scan set, though this may reduce the overall pruning effectiveness of the $\land$-expression.
Henceforth, pruning for filters $p3$ and $p4$ in \Cref{fig:pruner-tree} can be stopped if they prove ineffective.
In contrast, if the filter is part of an $\lor$-expression, removing that filter---thus assuming that every subsequent micro-partition \emph{might} contain qualifying rows---would effectively stop further pruning.
The removed branch from the $\lor$-expression would mark every micro-partition as potentially containing relevant rows and hence the $\lor$-expression itself becomes ineffective.
The $\lor$-expression itself may even be removed instead, which might recursively affect higher levels, potentially removing all pruning filters.
For instance, if $p1$ in \Cref{fig:pruner-tree} is removed, the left sub-tree would lose its ability to prune any micro-partitions.
Since the root node is also an $\lor$-expression, the entire pruning tree would be rendered ineffective.
To avoid this situation, only filters below an $\land$-expression may be removed.

If pruning for a filter is halted, the filter itself is still retained for query execution. Additionally, if pruning for a slow filter is stopped during query optimization, pruning might still be deferred to the highly parallel query execution stage.%

\subsection{Impact of Filter Pruning}\label{s:filter-pruning-impact}

\begin{figure}[htb]
    \centering
    \includegraphics[width=\linewidth]{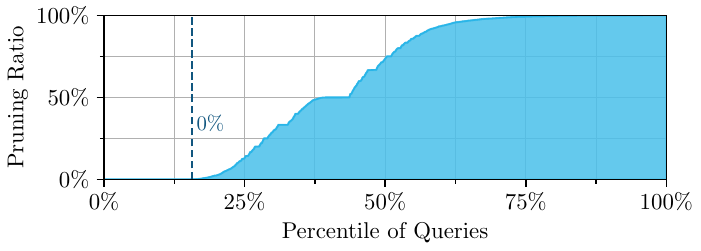}
    \caption{Impact of filter pruning on a representative random sample of \texttt{SELECT} queries that had at least one predicate. The sample is across all customers from Nov 5\textsuperscript{th} to Nov 7\textsuperscript{th} 2024.}
    \label{fig:filter-pruning-impact}
\end{figure}

In Snowflake, filter pruning is executed as the first of all pruning techniques and therefore its pruning ratio is independent of the pruning results of other techniques.
Further, filter pruning is generally applicable to all queries that contain at least one table scan with filters that allow pruning.
    
\Cref{fig:filter-pruning-impact} shows the impact of filter pruning.
The pruning ratio should be understood relative to the total number of micro-partitions that are to be processed by a query, including micro-partitions from table scans without any filters.
We chose to do so to give insights into the effect of pruning across the whole query and not just single table scans.

Overall, filter pruning is extremely effective, pruning at least around 90\% of partitions for 36\% of queries.
In some cases, filter pruning removes the whole scan set of a table scan, allowing additional optimizations such as sub-tree elimination in queries.
As can be seen in \Cref{fig:filter-pruning-impact}, around 27\% of queries have filters that allow for pruning but do not see a reduction in scan set size. This can have two possible reasons. Either, all partitions indeed contain qualifying rows or the data is poorly distributed with wide min/max ranges.

\section{Pruning for LIMIT Queries}
\label{s:limit-pruning}

\texttt{LIMIT} queries are common for BI workloads and frequently used during exploratory analysis of unfamiliar datasets~\cite{hanrahan2012}.
They account for a significant portion of queries executed on Snowflake (see \Cref{tab:limit-workload}).
Because such queries return only a subset of the qualifying rows, they show a significant potential for pruning.

Most existing database systems simply execute the entire query and halt query processing when the specified \texttt{LIMIT} has been reached.
Kim\etal~\cite{Kim2016} proposed the use of auxiliary index structures, called \emph{density maps}, together with specialized retrieval algorithms that prioritize processing "dense" partitions first.
These density maps are similar to per-partition histograms.
While their approach is elegant, our approach is globally IO-optimal for supported queries as it reads only the minimal number of files required. Further, our approach can be implemented with minimal modifications using only existing min/max metadata. 
However, it does come with the trade-off that our approach has stronger prerequisites and therefore supports fewer queries.

\subsection{Approach}

Continuing the example with the animal observation post, the next step a data analyst might take is getting an overview over the tracking data of alpine animals.
To do this, they issue the following \texttt{LIMIT} query on a table consisting of four micro-partitions as depicted in \Cref{fig:fmf}\footnote{Column \texttt{S} contains the height in [cm] for each animal, using realistic values.}.
In other scenarios, a cybersecurity expert might investigate a few connections from a specific IP address, a dashboard tool might automatically append a default \texttt{LIMIT} to all queries, or a data scientist might apply a \texttt{LIMIT} to source tables as a quick sampling method to reduce response times during model development.

\begin{lstlisting}[label=l:limit-example,language=SQL]
SELECT * FROM tracking_data
WHERE species LIKE 'Alpine%' AND s >= 50
LIMIT 3;
\end{lstlisting}

\noindent Applying filter pruning based on the query's predicate still requires scanning three micro-partitions. If we happen to scan partitions 2 and 4 first, the \texttt{LIMIT} would only be reached once the last partition is processed.
Ideally, we would identify partition 3 during query compilation as sufficient, allowing us to process only that micro-partition.
If the \texttt{LIMIT} exceeds the number of rows in partition 3, the ideal approach would be to process the micro-partitions in descending order of the number of qualifying rows to reduce the number of partitions that need to be scanned, similar to the density-map-based approach by Kim\etal~\cite{Kim2016}. However, knowing the number of qualifying rows in a partition is not always possible.

\colorlet{match}{snowflakeblue}
\colorlet{partialmatch}{snowflakelightblue}
\def\minmax#1#2{\;\rlap{\textsuperscript{\tiny$\geq#1$}}\textsubscript{\tiny$\leq#2$}}

\begin{figure}[b]
    \centering
    \tikzstyle{label} = [outer sep=0pt, inner sep=0pt, font=\normalsize\ttfamily]
    \begin{tikzpicture}[]
        \def\colwidthL{1.7cm}
        \def\colwidthR{.6cm}
        \node [shape=rectangle,align=center](table1) at (-1.9,1.4) {
            Partition 1\\
            \begin{tabular}{|C{\colwidthL}|C{\colwidthR}|} \hline
                \texttt{SPECIES}\minmax{\text{'B...'}}{\text{'S...'}} & \texttt{S}\minmax{7}{133}  \\ \hline
                Snow Vole & 7 \\
                Brown Bear & \cellcolor{partialmatch}133 \\
                Gray Wolf & \cellcolor{partialmatch}82 \\
                \hline
            \end{tabular}
        };
        \node [shape=rectangle,align=center](table2) at (1.9,1.4) {
            Partition 2\\
            \begin{tabular}{|C{\colwidthL}|C{\colwidthR}|} \hline
                \texttt{SPECIES}\minmax{\text{'A...'}}{\text{'R...'}} & \texttt{S}\minmax{6}{70}  \\ \hline
                Lynx & \cellcolor{partialmatch}71 \\
                Red Fox & 40 \\
                \cellcolor{partialmatch}Alpine Bat & 6 \\
                \hline
            \end{tabular}
        };
        \node [shape=rectangle,align=center](table3) at (-1.9,-1.4) {
            Partition 3\\
            \begin{tabular}{|C{\colwidthL}|C{\colwidthR}|} \hline
                \texttt{SPECIES}\minmax{\text{'A...'}}{\text{'A...'}} & \texttt{S}\minmax{76}{101}  \\ \hline
                \cellcolor{match}Alpine Ibex & \cellcolor{match}101 \\
                \cellcolor{match}Alpine Goat & \cellcolor{match}76 \\
                \cellcolor{match}Alpine Sheep & \cellcolor{match}83 \\
                \hline
            \end{tabular}
        };
        \node [shape=rectangle,align=center](table4) at (1.9,-1.4) {
            Partition 4\\
            \begin{tabular}{|C{\colwidthL}|C{\colwidthR}|} \hline
                \texttt{SPECIES}\minmax{\text{'A...'}}{\text{'P...'}} & \texttt{S}\minmax{4}{51}  \\ \hline
                Europ. Mole & 4 \\
                Polecat & 16 \\
                \cellcolor{match}Alpine Ibex & \cellcolor{match}97 \\
                \hline
            \end{tabular}
        };

        \node[draw, dashed, rounded corners, fit=(table1)](NotMatched) {};
        \node[left=0pt of NotMatched.west, rotate=90, anchor=south] {Not Matching};
        
        \node[draw, dashed, rounded corners, fit=(table3)](FullyMatched) {};
        \node[left=0pt of FullyMatched.west, rotate=90, anchor=south] {Fully Matching};
        
        \draw[dashed, rounded corners, name=PartiallyMatching] 
            ($(table2.north west) + (-0.1,0.1)$) -- 
            ($(table2.north east) + (0.1,0.1)$) coordinate (TopRight) -- 
            ($(table4.south east) + (0.1,-0.3)$) coordinate (BottomRight) -- 
            ($(table3.south west) + (-0.8,-0.3)$) -- 
            ($(table3.north west) + (-0.8,0.3)$) -- 
            ($(table4.north west) + (-0.1,0.3)$) -- 
            cycle;
        \coordinate (CenterRight) at ($(TopRight)!0.5!(BottomRight)$);
        \node[right=0pt of CenterRight.east, rotate=-90, anchor=south] {Partially Matching};
    \end{tikzpicture}
    \caption{Example of partially and fully matching partitions for a query with predicate \lstinline[keepspaces=true]|species LIKE 'Alpine\%' AND s >= 50| on a table with four micro-partitions. Metadata for \texttt{SPECIES} is shortened.  Colors highlight which rows/cells (partially) match the predicate.}
    \label{fig:fmf}
\end{figure}

To achieve this, we introduce a third category of micro-partitions: In addition to \textit{not-matching} partitions---those pruned away by filter pruning---and \textit{partially-matching} partitions---those that \emph{might} contain qualifying rows and are therefore retained in the scan---we define \textit{fully-matching} partitions, where we can guarantee that all rows of that partition qualify all query predicates.
By definition, \textit{fully-matching} partitions are a subset of \textit{partially-matching} partitions.
A similar idea called \textit{subsumed partitions}, although used in a different context, was proposed by Ding\etal~\cite{Ding2022}.

Given the definition of fully-matching partitions, we propose a pruning algorithm as follows: if the total row count across all fully-matching partitions meets or exceeds the threshold defined by \textit{k} in the LIMIT clause, the scan set is constructed to include only the minimum number of fully-matching partitions required to satisfy 
\textit{k}.
Conversely, if the row count of fully-matching partitions is less than 
\textit{k}, or if no fully-matching partitions exist, \texttt{LIMIT}-pruning is not possible.
Nevertheless, starting the table scan with fully-matching partitions promises faster query execution times.

\begin{figure}[tb]
    \centering
    \includegraphics[width=\linewidth]{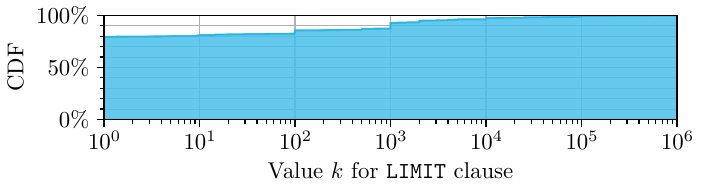}
    \caption{CDF of the distribution of $k$ in \texttt{LIMIT} queries (only $k>0$). If the query contained an \texttt{OFFSET}, the value for the offset is included. 97\% of queries have $k\leq10,000$ and 99.9\% have $k\leq2,000,000$. Queries are based on a representative random sample of queries with a \texttt{LIMIT} clause across all customers from Nov 5\textsuperscript{th} to Nov 7\textsuperscript{th} 2024.}
    \label{fig:limit-k-dist}
\end{figure}

Although our approach does not enable processing micro-par\-ti\-tions strictly in the order of their number of qualifying rows, in practice, the substantial size of partitions means that identifying one or two fully-matching partitions is typically sufficient to produce the query result.
Moreover, the value of $k$ in \texttt{LIMIT} queries is typically small\footnote{Some BI-tools issue queries with \texttt{LIMIT 0} appended to receive schema information of the output before sending the actual query.}, with most queries having $k=0$ or $k=1$ (\Cref{fig:limit-k-dist}). This suggests that optimizing for small values of $k$ is practical.

\begin{table}[tb]
    \centering
    \caption{Relative frequency of different types of \texttt{LIMIT} queries out of all \texttt{SELECT} queries in Snowflake over the course of three days from Nov 5\textsuperscript{th} to Nov 7\textsuperscript{th} 2024, based on pattern-matching on SQL texts.}
    \label{tab:limit-workload}
    \begin{tabular}{l c}
        \toprule
        \textbf{Type}   & \textbf{Percentage} \\
        \midrule
        \texttt{LIMIT} queries                                    & 2.60\percent \\
        \hspace{0.4cm}\texttt{LIMIT} without predicate            & 0.37\percent \\
        \hspace{0.4cm}\texttt{LIMIT} with predicate               & 2.23\percent \\
        Top-k queries                                             & 5.55\percent \\
        \hspace{0.4cm}\texttt{ORDER BY x LIMIT k}              & 4.47\percent \\
        \hspace{0.4cm}\texttt{GROUP BY x ORDER BY x LIMIT k}      & 0.12\percent \\
        \hspace{0.4cm}\texttt{GROUP BY y ORDER BY agg(x) LIMIT k} & 0.96\percent \\
        \bottomrule
    \end{tabular}
\end{table}

\subsection{Identifying Fully-Matching Partitions}\label{s:fmf}
A fully-matching partition is defined as a micro-partition only containing rows satisfying all query predicates.
In fact, the filter pruning process can be extended to identify such partitions by including a second pass with the inverted predicate, but without modifying the scan set during this step.

For instance, in the scenario illustrated in \Cref{fig:fmf}, the first pruning pass uses the predicate \lstinline[keepspaces=true]|species LIKE 'Alpine%' AND s >= 50|.
This removes partition 1 from the scan set based on its metadata for column \texttt{SPECIES}, leaving partitions 2, 3, and 4 marked as partially-matching.
In the second pass, the inverted predicate \lstinline[keepspaces=true]|species NOT LIKE 'Alpine%' OR s < 50| is applied, under which partition 3 is identified as not-matching.
But instead of excluding partition 3 from the scan set, this step confirms that it contains no rows failing the inverted predicate. Consequently, partition 3 is marked as fully-matching and retained in the scan set.
Trivially, if a query does not contain predicates, all partitions are considered fully-matching.
For queries with predicates, the existence of fully-matching partitions is not guaranteed.
Nonetheless, if such partitions do exist, the outlined procedure will accurately identify them.

\subsection{Considerations}
The first step of pruning LIMIT queries is to determine which table scan's scan set should be pruned.
This requires \enquote{pushing down} the LIMIT information within the compiler, from the LIMIT operator to the table scans.
Generally, operators that reduce the number of rows prevent this pushdown.
This includes, \eg filters where identifying fully-matching partitions is not possible, aggregations, joins, \etc
A noteworthy exception to this is propagating the \texttt{LIMIT} information through the build side of a (LEFT) OUTER JOIN, which still yields a correct query result.
When the LIMIT information reaches a table scan, the corresponding scan set can be pruned accordingly.

\subsection{Impact of LIMIT Pruning}\label{s:limit-impact}

\Cref{tab:limit-pruning-impact} shows the effect of pruning for \texttt{LIMIT} queries.
The majority of queries already has a minimal scan set of just a single micro-partition after filter pruning, leaving no opportunity for additional \texttt{LIMIT} pruning.
Out of the remaining queries we observe that there is a significant number of queries \emph{with predicate} that have an unsupported shape or without \emph{fully-matching} partitions that could be used for pruning. This shows that there is still a significant opportunity to improve pruning for \texttt{LIMIT} queries when no \emph{fully-matching} partitions are found.
Conversely, if the prerequisites for pruning are met, doing so reduces the number of partitions significantly, mostly to just 1 partition, regardless of the original number of partitions. The cases where there are more than 1 partition left are due to large $k$ in the \texttt{LIMIT} clause and still result in the optimal number of partitions.
All in all, we conclude that there are still many queries that we can not apply this pruning technique to---but if we can, pruning results in a drastic reduction of the scan set.

Pruning for \texttt{LIMIT} queries should be performed \emph{after} filter pruning, as a simple extension to filter pruning can provide the necessary information about fully-matching partitions.

It should be noted, however, that for \texttt{LIMIT} queries, the pruning ratio does not always result in a corresponding runtime or even IO improvement. The reason for this is that the query engine would halt execution after enough rows have been processed and most micro-partitions would not have been scanned anyways.
However, there is a catch for parallel query execution: if no pruning is applied, the work might be distributed across $n$ machines, each required to scan up to $\lceil k/n \rceil$ rows. This usually means that the query engine reads at least $n$ partitions, even though $1$ might have been enough.

\begin{table}[htb]
    \centering
    \caption{Breakdown of \texttt{LIMIT} pruning applicability, showing the percentage of \texttt{LIMIT} queries falling into each category. We further make a distinction between \texttt{LIMIT} queries with and without predicates. Queries are based on a representative random sample of eligible queries across all customers from Nov 5\textsuperscript{th} to Nov 7\textsuperscript{th} 2024.}
    \label{tab:limit-pruning-impact}
    \begin{tabular}{p{3.4cm} C{1.34cm} C{1.34cm} C{1.0cm}}
        \toprule
        \textbf{Queries with...}   & \textbf{Without Predicate} & \textbf{With Predicate} & \textbf{Overall}\\
        \midrule
        already minimal scan set & 79.60\% & 61.65\% & 64.22\% \\
        unsupported shapes & 1.74\% & 36.23\% & 31.28\% \\
        pruning to $= 1$ partition & 16.58\% & 1.71\% & 3.85\% \\
        pruning to $>1$ partitions & 1.54\% & 0.01\% & 0.23\% \\
        \bottomrule
    \end{tabular}
\end{table}

\section{Pruning for Top-K Queries}
\label{s:topk-pruning}

Snowflake employs a specialized pruning technique for top-k que\-ries during execution, complementing other pruning mechanisms.
Similar to \texttt{LIMIT} queries, BI workloads often contain a significant number of top-k queries.
Notably, 5.55\% of all \texttt{SELECT} queries executed on Snowflake fall into this category.
\Cref{tab:limit-workload} presents a breakdown of different types of \texttt{LIMIT} queries, highlighting their relative frequencies within Snowflake over a one-week period.
Top\nobreakdash-k queries are characterized by the use of an \texttt{ORDER BY} clause in conjunction with a \texttt{LIMIT} clause.
This means that---logically---the query engine first needs to perform sorting on all qualifying tuples before applying the \texttt{LIMIT}.
However, few systems actually do this, but instead employ a heap-based approach, as we briefly describe in the next paragraph.

To illustrate this concept, we build upon the last example. A good animal observation post should have a high chance of seeing animals of a certain kind, so the data analyst might try to increase the number of animal sightings by ordering by \texttt{num\_sightings}:
\begin{lstlisting}[label=l:topk-example,language=SQL]
SELECT * FROM tracking_data
WHERE species LIKE 'Alpine%' AND s >= 50
ORDER BY num_sightings DESC LIMIT 3;
\end{lstlisting}
Similarly, a BI-dashboard might show the top-10 users or a threat-detection tool might identify recent log-in attempts.
Note that the \texttt{ORDER BY} expression might additionally be wrapped in a function.

\medskip
\paragraphnospace{Standard Heap-Based Approach.}
A widely used method~(\eg\cite{raasveldt2019duckdb, stonebraker1986design}) for processing top-k queries involves maintaining a min-heap (for queries with \texttt{DESC} ordering) that holds up to \textit{k} elements.
Qualifying tuples are progressively inserted into the heap as the table is scanned, and the heap’s contents are returned as the final query result. 
This method scales linearly with the size of the table, becoming expensive for large tables in data analytics systems, as it requires scanning the entire table.

\medskip
\paragraphnospace{The Theoretically Optimal Solution.}
Ideally, if the table is physically sorted by the \texttt{ORDER BY} key, the query engine should only need to scan the number of micro-partitions required to locate the first \textit{k} qualifying rows.
For instance, in the absence of predicates and assuming each micro-partition contains more than three rows, the engine would only need to scan a single partition if the table is sorted by \texttt{num\_sightings}.
Conversely, if the data is not sorted by \texttt{num\_sightings}, the query engine should---ideally---scan only the micro-partitions containing the three highest values in the \texttt{num\_sightings} column.
In this example, this would involve reading at most three micro-partitions.

\subsection{Existing Research}\label{s:topk-related-work}

A substantial body of research in the Information Retrieval (IR) domain focuses on optimizing the deterministic retrieval of top-k elements when joining multiple \enquote{posting lists} containing \texttt{(doc\_id, attributes)} pairs~\cite{Ilyas2008}.
In this context, the final score for a document is computed using a scoring function that aggregates the individual attributes distributed across these lists.

\medskip
\paragraphnospace{The Threshold Algorithm (TA).}
When these lists are sorted by the scoring attribute, the challenge is to efficiently find and communicate a halting point when iterating through the lists.
Once a halting point is found, it ensures that no additional tuples can qualify for the result.
This traversal technique is often referred to as Term-At-A-Time (TAAT).
Fagin\etal~\cite{Fagin2001} proposed the Threshold~Algorithm~(TA) for this kind of problem which has been extensively studied and refined ~\cite{Cao2004,Pang2010}.

\medskip
\paragraphnospace{The WAND Algorithm.}
In contrast, when the lists are sorted by \textit{doc\_id}, the traversal strategy known as Document-At-A-Time (DAAT) resembles the relational model where tables are split into separate columns.
This data model aligns closely with processing top-k queries in relational databases, making it particularly relevant.

Broder \etal~\cite{Broder2003} introduced the WAND algorithm, which employs a continuously updated \emph{threshold value} to skip documents with scores lower than the threshold.
This threshold, representing the $k$-th largest score encountered thus far, is conceptually similar to the pruning boundary in our approach.

Around the same time, Chakrabarti\etal~\cite{Chakrabarti2011} and Ding \etal~\cite{Ding2011} independently proposed extensions to the WAND algorithm that introduced the concept of a \textit{block-max} value, enabling skipping of larger segments of data.
\cite{Chakrabarti2011} further suggested to adjust the processing order of blocks based on their \textit{block-max} value.
However, both works lacked ideas for initializing the threshold value besides setting it to zero.
\cite{Ding2011} initially assumed fixed-sized blocks of arbitrary size, a design later enhanced in \cite{Mallia2017} to use variable-sized blocks.
In contrast, Chakrabarti\etal~\cite{Chakrabarti2011} had strict requirements for splitting lists into blocks, resulting in very small partitions. Here, the \textit{score} value is allowed to change only at block boundaries, meaning each block must contain a uniform \textit{score} value. This assumption is impractical for large-scale data processing as it necessitates physical reorganization data. Subsequent research expanded upon the concept of data skipping using \textit{block-max} indexes~\cite{Rojas2013,Dimopoulos2013,Dimopoulos2013.2,Bortnikov2017,Khattab2020}.
All of the aforementioned work lies within the IR space and does not explore ways to fully integrate these approaches into a relational query engine, besides some initial ideas presented in~\cite{Fagin2001}.

\medskip
\paragraphnospace{Provenance Sketches.}
Another notable approach for processing top-k queries is presented by Niu \etal~\cite{Niu2021}, who utilize per-partition \textit{provenance sketches} to enhance data pruning for such queries.
Notably, their method supports data skipping even when the \texttt{ORDER BY} clause involves an aggregated column.
However, maintaining these sketches efficiently under data updates poses significant challenges.

\begin{figure*}[t]
    \tikzstyle{label} = [outer sep=0pt, inner sep=0pt, font=\normalsize\ttfamily]
    \tikzset{
        box/.style={text centered, inner sep=3, align=center,font=\small}
    }
    \def\figheight{3.5cm}
    \begin{subfigure}[t]{0.24\textwidth}
        \begin{minipage}[t][\figheight][t]{\textwidth} %
            \centering
            \begin{tikzpicture}[baseline={(current bounding box.north)}]
                \def\xdir{10}
                \def\ydir{2}
                \node (out) [box] {...};
                \node (topk) [box, below=\ydir*1mm of out] {\huge\sort};
                    \node [anchor=north west,font=\scriptsize\ttfamily,outer sep=0pt, inner sep=0pt] at ($(topk.south east)-(.2em,-.5em)$) {by=c1, limit=k};
                \node (proj) [box, below=\ydir*1mm of topk] {\Large\proj};
                    \node [anchor=north west,font=\scriptsize\ttfamily,outer sep=0pt, inner sep=0pt] at ($(proj.south east)-(.2em,-.5em)$) {c1,\dots};
                \node (tablescan) [box, below=\ydir*1mm of proj] {\Large$\mathcal{R}$};
                \draw[thick] (out) -- (topk);
                \draw[thick] (topk) -- (proj);
                \draw[thick] (proj) -- (tablescan);
                \draw[-latex,dashed] ($(topk.south east)-(-.5cm, .2em)$) to[out=-30,in=20,distance=.6cm] (tablescan.east);
            \end{tikzpicture}
        \end{minipage}
        \caption{TopK above Table Scan}
        \label{fig:sub:topk-above-table-scan}
    \end{subfigure}
    \begin{subfigure}[t]{0.24\textwidth}
        \begin{minipage}[t][\figheight][t]{\textwidth} %
            \centering
            \begin{tikzpicture}[baseline={(current bounding box.north)}]
                \def\xdir{10}
                \def\ydir{2}
                \node (out) [box] {...};
                \node (topk) [box, below=\ydir*1mm of out] {\huge\sort};
                    \node [anchor=north west,font=\scriptsize\ttfamily,outer sep=0pt, inner sep=0pt] at ($(topk.south east)-(.2em,-.5em)$) {by=c1, limit=k};
                \node (join) [box, below=\ydir*.75mm of topk] {\huge\join};
                \node (projright) [box, below=\ydir*1mm of join, xshift=\xdir*1mm] {\Large\proj};
                    \node [anchor=north west,font=\scriptsize\ttfamily,outer sep=0pt, inner sep=0pt] at ($(projright.south east)-(.2em,-.5em)$) {c1,\dots};
                \node (tablescanright) [box, below=\ydir*1mm of projright] {\Large$\mathcal{S}$};
                \node (projleft) [box, below=\ydir*1mm of join, xshift=\xdir*-1mm] {\Large\proj};
                \node (tablescanleft) [box, below=\ydir*1mm of projleft] {\Large$\mathcal{R}$};
                \draw[thick] (out) -- (topk);
                \draw[thick] (topk) -- (join);
                \draw[thick] (join) -- (projleft);
                \draw[thick] (projleft) -- (tablescanleft);
                \draw[thick] (join) -- (projright);
                \draw[thick] (projright) -- (tablescanright);
                \draw[-latex,dashed] ($(topk.south east)-(-.5cm, .2em)$) to[out=-30,in=20,distance=1.3cm] (tablescanright.east);
            \end{tikzpicture}
        \end{minipage}
        \caption{TopK Join Probe Side}
        \label{fig:sub:topk-join-probe-side}
    \end{subfigure}
    \begin{subfigure}[t]{0.24\textwidth}
        \begin{minipage}[t][\figheight][t]{\textwidth} %
            \centering
            \begin{tikzpicture}[baseline={(current bounding box.north)}]
                \def\xdir{10}
                \def\ydir{2}
                \node (out) [box] {...};
                \node (topk) [box, below=\ydir*1mm of out] {\huge\sort};
                    \node [anchor=north west,font=\scriptsize\ttfamily,outer sep=0pt, inner sep=0pt] at ($(topk.south east)-(.2em,-.5em)$) {by=c1, limit=k};
                \node (join) [box, below=\ydir*.75mm of topk] {\huge\louterjoin};
                \node (projright) [box, below=\ydir*1mm of join, xshift=\xdir*1mm] {\Large\proj};
                \node (tablescanright) [box, below=\ydir*1mm of projright] {\Large$\mathcal{R}$};
                \node (topkleft) [box, below=\ydir*1mm of join, xshift=\xdir*-1mm] {\huge\sort};
                    \node [anchor=north west,font=\scriptsize\ttfamily,outer sep=0pt, inner sep=0pt] at ($(topkleft.south east)-(.2em,-.5em)$) {by=c1, limit=k};
                \node (projleft) [box, below=\ydir*1mm of topkleft] {\Large\proj};
                    \node [anchor=north west,font=\scriptsize\ttfamily,outer sep=0pt, inner sep=0pt] at ($(projleft.south east)-(.2em,-.5em)$) {c1,\dots};
                \node (tablescanleft) [box, below=\ydir*1mm of projleft] {\Large$\mathcal{S}$};
                \draw[thick] (out) -- (topk);
                \draw[thick] (topk) -- (join);
                \draw[thick] (join) -- (topkleft);
                \draw[thick] (topkleft) -- (projleft);
                \draw[thick] (projleft) -- (tablescanleft);
                \draw[thick] (join) -- (projright);
                \draw[thick] (projright) -- (tablescanright);
                \draw[thick,double] (topk) -- (topkleft);
                \draw[-latex,dashed] ($(topkleft.south east)-(-.5cm, .2em)$) to[out=-30,in=20,distance=.6cm] (tablescanleft.east);
            \end{tikzpicture}
        \end{minipage}
        \caption{TopK Outer Join Build}
        \label{fig:sub:topk-outer-join}
    \end{subfigure}
    \begin{subfigure}[t]{0.24\textwidth}
        \begin{minipage}[t][\figheight][t]{\textwidth} %
            \centering
            \begin{tikzpicture}[baseline={(current bounding box.north)}]
                \def\xdir{10}
                \def\ydir{2}
                \node (out) [box] {...};
                \node (topk) [box, below=\ydir*1mm of out] {\huge\sort};
                    \node [anchor=north west,font=\scriptsize\ttfamily,outer sep=0pt, inner sep=0pt] at ($(topk.south east)-(.2em,-.5em)$) {by=c1, limit=k};
                \node (agg) [box, below=\ydir*.75mm of topk] {\Large\group};
                    \node [anchor=north west,font=\scriptsize\ttfamily,outer sep=0pt, inner sep=0pt] at ($(agg.south east)-(.2em,-.5em)$) {by=c1, limit=k};
                \node (proj) [box, below=\ydir*1mm of agg] {\Large\proj};
                    \node [anchor=north west,font=\scriptsize\ttfamily,outer sep=0pt, inner sep=0pt] at ($(proj.south east)-(.2em,-.5em)$) {c1,\dots};
                \node (tablescan) [box, below=\ydir*1mm of proj] {\Large$\mathcal{R}$};
                \draw[thick] (out) -- (topk);
                \draw[thick] (topk) -- (agg);
                \draw[thick] (agg) -- (proj);
                \draw[thick] (proj) -- (tablescan);
                \draw[-latex,dashed] ($(agg.south east)-(-.5cm, .2em)$) to[out=-30,in=20,distance=.6cm] (tablescan.east);
            \end{tikzpicture}
        \end{minipage}
        \caption{TopK Aggregation}
        \label{fig:sub:topk-aggregation}
    \end{subfigure}

    \caption{Supported query plans for top-k pruning where \sort\ is used to represent a TopK operator.}
    \label{fig:topk-supported-operators}
\end{figure*}
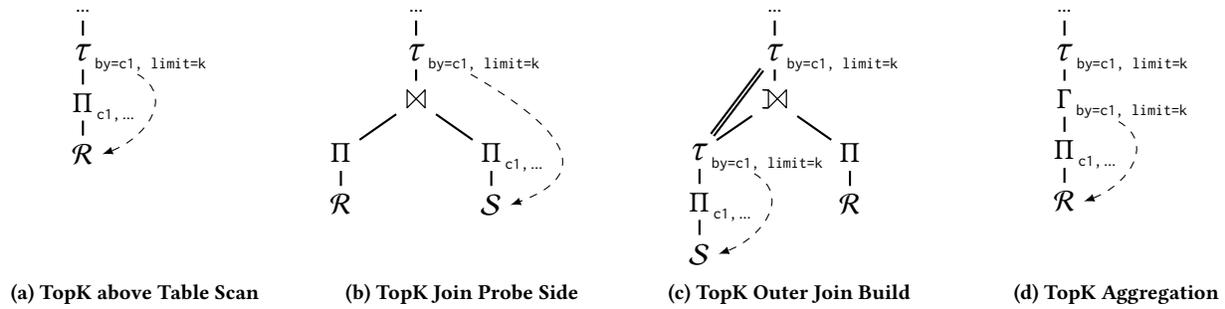

\subsection{Approach}\label{s:topk-approach}

For simplicity, the following section assumes \textit{descending} ordering of a top-k query.
Similar to \cite{Chakrabarti2011} and \cite{Ding2011}, we extend the heap-based approach and leverage the smallest element in the top-k heap---referred to as the \emph{boundary value}---for partition pruning.
Initially, no boundary value exists, as the heap is empty or contains fewer than $k$ rows.
Once $k$ rows have been inserted, the boundary value is updated with each new insertion.
Snowflake's flexible execution engine supports passing this updated boundary value to the table scan, enabling effective pruning.
Before scanning a micro-partition, its metadata is compared against the boundary value.
If the partition’s maximum value for the \texttt{ORDER BY} column is smaller than the boundary value, the micro-partition can be safely skipped, as none of its rows would qualify for the top-k heap.
As the query progresses and the heap approaches the final result, the boundary value becomes tighter, allowing for increased partition pruning.

\medskip
\paragraphnospace{Identifying Supported Table Scans.}
\noindent In general, this pruning technique can only be applied if the TopK operator is part of the same pipeline as the table scan that produces the \texttt{ORDER BY} column, meaning that the TopK operator has to sit \enquote{on top} of the table scan without any pipeline breakers in between.
Specifically, this means that there might be a filter between the table scan and the TopK operator, reducing the number of rows that make it into the top-k heap (see \Cref{fig:sub:topk-above-table-scan}).
Still, partition pruning is possible in this case because the boundary value is created based on the rows that make it into the heap.
Further, a TopK operator might also be located on top of a \texttt{JOIN} and prune micro-partitions coming from the \emph{probe side} of that \texttt{JOIN} if the \texttt{ORDER BY} column is produced by that side (see \Cref{fig:sub:topk-join-probe-side}).
Additionally, for a \texttt{(LEFT) OUTER JOIN} and for the case that the \texttt{ORDER BY} column is produced by the build side, the TopK operator can be replicated to the \emph{build side} of the \texttt{JOIN}, effectively supporting partition pruning on the build side in such cases (see \Cref{fig:sub:topk-above-table-scan}).
This is because we can guarantee that all $k$ rows from the build side will be forwarded beyond the \texttt{JOIN}.

\medskip
\paragraphnospace{Supporting Aggregations.}
In certain cases, it is possible to prune micro-partitions via the TopK operator when there is an aggregation, \ie a \texttt{GROUP BY}, between the TopK node and the table scan (see \Cref{fig:sub:topk-aggregation}).
If the ordering is not performed on an aggregate column, \ie the \texttt{ORDER BY} columns are a subset of the \texttt{GROUP BY} keys, pruning is possible.
However, this requires changes to the \texttt{GROUP BY} operator to maintain its own top-k heap.

\subsection{Processing Order of Data Partitions}\label{s:topk-scanset-sorting}

The effectiveness of runtime partition pruning depends largely on identifying a tight threshold early and the extent of overlap in micro-partition min/max values.
Since our work deliberately assumes we do not have control over the physical layout of data, we cannot influence by how much micro-partitions overlap.
Still, the min/max information can be used to process micro-partitions in an order that promises a good threshold value early on.

While the basic idea is simple, there are several considerations when applying this approach in a large-scale, cloud-based data analytics system.
For top-k queries with \texttt{DESC} ordering, the micro-partitions would be sorted by their \textit{max} values, starting processing with the partition that has the largest \textit{max} value.
This strategy prioritizes scanning partitions with larger values first, enabling the construction of a top-k heap that closely approximates the final query result early in the processing phase.
Subsequently scanned partitions then refine the result further if their rows are not fully filtered out by predicates.

However, a naive sorting approach can negatively impact performance. 
Sorting, especially for datasets with millions of micro-partitions, incurs a non-negligible computational overhead.
Additionally, for queries with highly selective filters, naive sorting might lead to a worse processing order of partitions.
In such a case, naive sorting might accidentally de-prioritize scanning micro-partitions that actually contain matching rows, potentially leading to a situation where lots of partitions with \enquote{large} values are processed that do not contain any qualifying tuples.
This, in turn, means that more micro-partitions need to be processed until the top-k heap is fully populated and pruning can start.
Unfortunately, sorting based on min/max metadata only cannot guarantee the presence of qualifying rows.
Thus, a sorting strategy needs to account for that.

\noindent We evaluated the following sorting strategies:
\begin{enumerate}
    \item \textbf{None/random}. No explicit sorting is applied. Instead, a random processing order is used.
    \item \textbf{Full sort}. A standard sorting algorithm sorting all micro-partitions by their min/max values.
\end{enumerate}

\begin{figure}[htb]
    \centering
    \includegraphics[width=\linewidth]{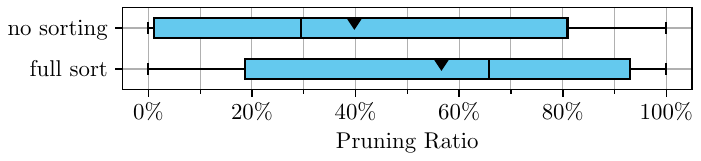}
    \caption{Influence of sorting on the pruning ratio for top-k queries on a representative random sample of eligible queries across all customers on a 7-day interval from Nov 13\textsuperscript{th} to Nov 19\textsuperscript{th} 2024. The sample contains only queries with a minimum runtime of 1 second when top-k pruning was disabled.}
    \label{fig:scanset-sorting}
\end{figure}

\noindent\Cref{fig:scanset-sorting} shows the influence of sorting on partition pruning together with a baseline of \enquote{no sorting}.
Overall, sorting significantly improves the average pruning ratio compared to a random partition order.
This not only becomes visible in the improved median but also in the distribution tails, resulting in better worst-case performance. It demonstrates that sorting can often produce a tighter boundary value early, enabling more effective partition pruning.

\subsection{Upfront Initialization of Boundary Values}\label{s:topk-comp-time-boundary-value}
In our algorithm, pruning begins only after the top-k heap contains enough rows.
Early in query execution, any row can enter the heap, often resulting in a sub-optimal boundary value.
However, if the metadata includes details like the number of rows per micro-partition and whether a column contains null values, we can pre-compute an initial boundary value during query compilation, enabling pruning from the very first partition.

The approach again uses \emph{fully-matching} partitions, as described in \Cref{s:fmf}.
Given a set of fully-matching partitions, for \texttt{DESC}-ordering queries, the boundary can be initialized using either the $k$-th largest max-value of the \texttt{ORDER BY} column of all fully-matching partitions or by sorting these micro-partitions by their min-values of the \texttt{ORDER BY} column and selecting the min-value of the first micro-partition where the cumulative row count of all previous partitions meets or exceeds $k$, whichever yields a stricter boundary.
Typically, for highly overlapping micro-partitions, the $k$-th max-value is generally more effective, while for (partially) sorted tables, the largest min-value is often the better choice.

\subsection{Impact of Top-K Pruning}\label{s:topk-impact}

\begin{figure*}[t]
    \centering
    \begin{subfigure}[t]{0.32\linewidth}
        \includegraphics[width=\linewidth]{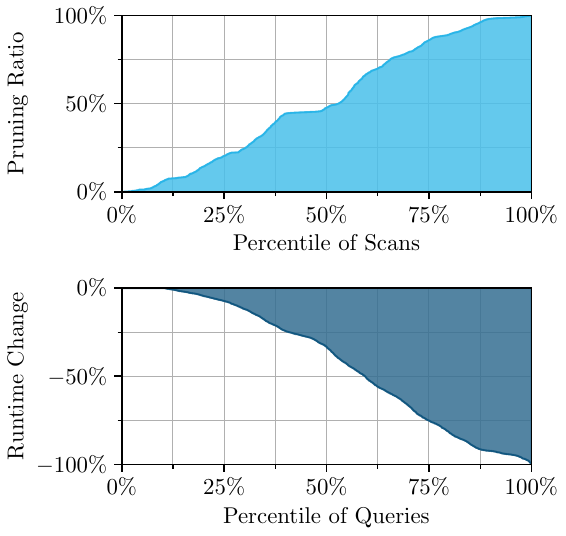}
        \caption{$1\,s < t \leq 10\,s$}
        \label{fig:sub:topk-impact-short}
    \end{subfigure}
    \begin{subfigure}[t]{0.32\linewidth}
        \includegraphics[width=\linewidth]{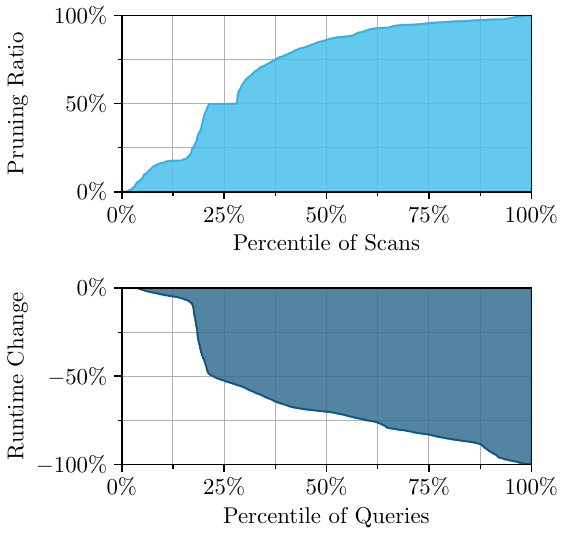}
        \caption{$10\,s < t \leq 60\,s$}
        \label{fig:sub:topk-impact-medium}
    \end{subfigure}
    \begin{subfigure}[t]{0.32\linewidth}
        \includegraphics[width=\linewidth]{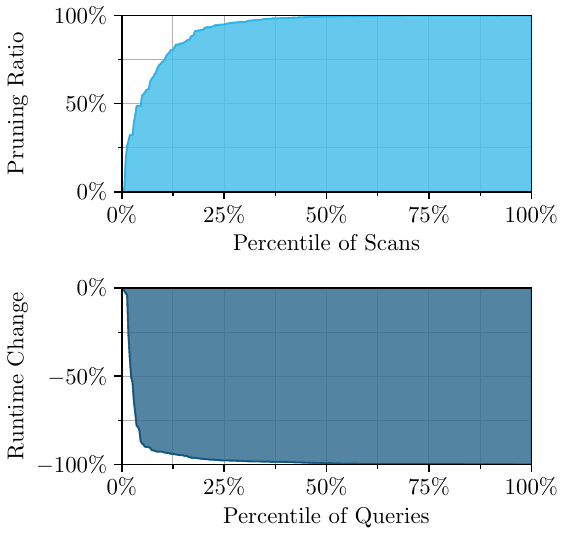}
        \caption{$t > 60\,s$}
        \label{fig:sub:topk-impact-long}
    \end{subfigure}

    \caption{The first row shows CDFs of the pruning ratio (higher is better) of top-k pruning on table-scan level where top-k pruning was successfully applied. The plots in the second row show CDFs of the relative runtime improvement of queries. Queries are randomly sampled from representative real-world workloads over a period of 7-days from Nov 13\textsuperscript{th} to Nov 19\textsuperscript{th} 2024, bucketed by query execution times. The execution time $t$ of a query was measured when top-k pruning was disabled.}
    \label{fig:topk-impact}
\end{figure*}

\revised[1] {
In Snowflake, top-k pruning is applied after filter pruning, \texttt{JOIN} pruning, and potentially \texttt{LIMIT} pruning, meaning its effectiveness depends on the combined outcome of these preceding techniques.
\Cref{fig:topk-impact} shows pruning ratios for table scans where top-k pruning was successfully applied and its effect on overall query runtimes.
The CDFs have similar distributions, indicating a strong correlation between pruning and runtime improvement.
We categorized queries into buckets based on their runtimes when top-k pruning was disabled, allowing us to observe its impact on both fast and slow queries.
Across all buckets, we observed query runtime improvements of more than 99.9\%.
Out of the queries in our sample, 4.8\% also applied \texttt{JOIN}-pruning.
Overall, the average pruning ratio of successfully applied top-k pruning is around 77\%.}

\section{Pruning for JOIN Queries}
\label{s:join-pruning}

Joins are some of the most time-consuming and compute-intensive operations in a database system, making it crucial to avoid unnecessary work during these operations.
This section explains how partition pruning can be used to improve join queries significantly.

Join pruning can be seen as a form of sideways information passing~\cite{ives2008sideways}.
Unlike row-level techniques like \emph{bloom-joins}~\cite{mackert1986r, mullin1990optimal}, join pruning in Snowflake operates at a coarser granularity, excluding entire micro-partitions from join processing on the probe side.

\subsection{Approach}

Continuing the example, we identify a suitable location for the animal observatory.
The goal is to select a site along an existing hiking trail on a mountain ridge that offers high chances of observing large, local wildlife:
\begin{lstlisting}[label=l:join-example,language=SQL]
SELECT * 
FROM trails t JOIN tracking_data d ON t.mountain = d.area
WHERE true
  AND IF(unit='feet', altit * 0.3048, altit) > 1500 
  AND name LIKE 'Marked-%-Ridge'
  AND species LIKE 'Alpine%' AND s >= 50
ORDER BY d.num_sightings DESC LIMIT 3;
\end{lstlisting}

\noindent The overarching goal of join pruning is to reduce the work done on the probe side of the join.
Work can be reduced at different levels.
(i)~Computational work can be reduced by not having to check if a probe side value is part of the build side's hash table.
(ii)~In addition, and most important for a cloud-based data processing system with decoupled compute and storage, we can reduce (network) I/O by avoiding to load probe side micro-partitions altogether.

\noindent The general idea of join pruning %
can be divided into four steps:
\begin{enumerate}
    \item Summarize the entirety of the build side values during the hash join's build phase.
    \item Ship the previously generated summary from the build side to the probe side.
    \item Match the summary against the join probe side's micro-partitions' min/max values.
    \item Prune those micro-partitions whose min/max values do not overlap with the summary.
\end{enumerate}

\noindent In our example, selective filters on the build side substantially reduce the \texttt{trails} table, creating pruning opportunities on the probe side.
Further, the probe side already underwent filter pruning
and top-k pruning can be applied during the join---resulting in three distinct pruning techniques being used on the \texttt{tracking\_data} table.

\medskip
\paragraphnospace{Summarizing Build-Side Values.}
Summarizing build side values is a trade-off between accuracy and the memory size of the employed data structure.
In a distributed setting, this summary needs to be sent to other workers over the network, which forbids excessive data sizes.
A very cheap but inaccurate value summary would only store the global min/max values of the build-side which comes with negligible storage overhead but low pruning potential.
Most of the time, a Bloom-filter~\cite{bloom1970space} or a similar data structure is used in sideways information passing to summarize build-side values.
The Bloom-filter is then used to skip probing the join's hash table for individual rows, which reduces CPU usage significantly.

The value summary technique employed by Snowflake allows pruning whole micro-partitions and strikes a balance between accuracy and storage cost: while spending only a small fraction of the build-side size for the summary, Snowflake does still manage to prune 79\% of micro-partitions of probe-side scans.
In practice, we observed probe-side scan sets being reduced by up to 99.99\%.
This trade-off demonstrates the probabilistic character of the approach.
For pruning micro-partitions of the probe-side, build-side value summaries are overlapped with probe-side partition min/max metadata to efficiently check whether a micro-partition might contain any joinable tuples.
Notably, this does not just reduce CPU usage by avoiding checking tuples against the hash table of the build-side, but also significantly reduces network I/O because micro-partitions are pruned before they are loaded from storage.

\subsection{Limitations}
The join pruning technique discussed in this section is targeted towards hash-based joins.
In addition, join pruning requires a storage model as explained in \Cref{s:snowflake-architecture} consisting of micro-partitions with per-column min/max metadata.
The join pruning technique presented is probabilistic, \ie it might miss pruning a micro-partition that could theoretically be pruned, but will never prune a partition that must not be pruned.

\subsection{Impact of Join Pruning}\label{s:join-pruning-impact}

\begin{figure}[tb]
    \centering
    \includegraphics[width=\linewidth]{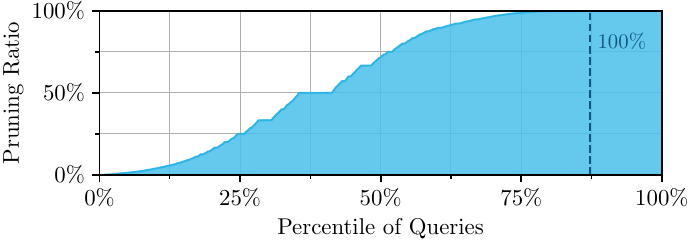}
    \caption{Impact of join pruning for \texttt{SELECT} queries that were able to successfully use join pruning. Based on a representative random sample of eligible queries across all customers from Nov 5\textsuperscript{th} to Nov 7\textsuperscript{th} 2024.}
    \label{fig:join-pruning-impact}
\end{figure}

Join pruning is executed after compile-time pruning and therefore its pruning ratio depends on filter pruning and---in theory---on \texttt{LIMIT} pruning. However, a \texttt{LIMIT} below a join rarely makes sense.
Filter pruning can incorporate filters from the build side to a limited extent only, as demonstrated by a technique known as \emph{data-induced predicates}~\cite{kandula2019pushing}.
Join pruning, on the other hand, has a full overview over the tuples on the build side and therefore is able to effectively prune micro-partitions on the probe side.
\Cref{fig:join-pruning-impact} depicts the average scan set reduction of probe-side tables resulting from join pruning.
Most notably, around 13\% of queries see a pruning ratio of 100\%, meaning that the probe-side scan does not need to be performed.
This might be caused by an empty build-side.
Further, we see that join pruning is generally very effective, with 50\% of queries seeing a scan set reduction of at least 72\%.

\section{The Pruning Flow in Snowflake}
\label{s:pruning-flow}

In the previous sections, we examined pruning techniques in isolation. Here, we briefly summarize their combined applicability.
\Cref{fig:pruning-flow} shows the share of queries with successful partition pruning by different technique combinations and the order in which Snowflake applies them. Filter pruning at execution time is omitted.
Note that this plot includes both DML and \texttt{SELECT} queries in Snowflake, reflecting a corresponding distribution of query types.
All percentages are relative to 100\%, though minimum heights have been applied to lines and boxes for readability.
As shown, most queries benefit from filter pruning. 
Note that the percentages indicate queries where at least one partition was pruned using a given technique---they do not imply, \eg that only 58.7\% of queries have a \texttt{WHERE} clause. A query may contain a \texttt{WHERE} clause but still be non-selective or non-prunable.

\newcommand{\round}[2]{%
    \pgfkeys{/pgf/number format/.cd, fixed, fixed zerofill, precision=#1}%
    \pgfmathprintnumber{#2}%
}

\begin{figure}[htb]
    \centering

\begin{tikzpicture}[x=0.97pt,y=0.82pt]
  \begin{sankeydiagram}[
    sankey tot length=82pt,%
    sankey tot quantity=100,%
    sankey min radius=15pt,%
    sankey arrow length=0,%
    sankey fill/.style={
      draw,line width=0pt,
      fill,
      snowflakelighterblue,
    },
    sankey draw/.style={
      draw=snowflakeblue,
      line width=0.3pt,
      line join=round,
    },
    bar label/.style={
        font={\scriptsize},
        align=center,
        anchor=center,
        white,
    },
    bar title/.style={
        font={\scriptsize},
        align=center,
        anchor=north,
        inner sep=1.5pt,
        black,
    },
    line label/.style={
        font={\tiny},
        align=center,
        anchor=center,
        inner sep=1pt,
        text=snowflakedarkblue,
    },
    background label/.style={
        font={\scriptsize},
        align=center,
        anchor=south east,
        inner sep=1pt,
        outer sep=1pt,
        fill=white,
        text=gray,
    },
    ]

\def\allToFilter{58.710865408398185}
\def\allToLimit{0.00017837097357255655}
\def\allToJoin{3.9837373237694784}
\def\allToTopk{0.0284798987804182}
\def\allToOut{37.27673899807835}
\def\filterToLimit{0.1963269849121939}
\def\filterToJoin{6.6975327726935445}
\def\filterToTopk{0.08306141669362051}
\def\filterToOut{51.73394423409883}
\def\limitToJoin{0.00011891398238170436}
\def\limitToTopk{0.0}
\def\limitToOut{0.19638644190338475}
\def\joinToTopk{0.0067780969957571486}
\def\joinToOut{10.674610913449648}
\def\topkToOut{0.11831941246979584}

\tikzmath{
\allFilter = \allToFilter;
\allLimit = \allToLimit + \filterToLimit;
\allJoin = \allToJoin + \filterToJoin + \limitToJoin;
\allTopk = \allToTopk + \filterToTopk + \joinToTopk;
}

\def\minLineHeight{1}
\def\lineArrowLength{0}
\tikzmath{
\allToFilterHeight = max(\allToFilter, \minLineHeight);
\allToLimitHeight = max(\allToLimit, \minLineHeight);
\allToJoinHeight = max(\allToJoin, \minLineHeight);
\allToTopkHeight = max(\allToTopk, \minLineHeight);
\filterToLimitHeight = max(\filterToLimit, \minLineHeight);
\filterToJoinHeight = max(\filterToJoin, \minLineHeight);
\filterToTopkHeight = max(\filterToTopk, \minLineHeight);
\limitToJoinHeight = max(\limitToJoin, \minLineHeight);
\limitToTopkHeight = max(\limitToTopk, \minLineHeight);
\joinToTopkHeight = max(\joinToTopk, \minLineHeight);
}
\tikzmath{
\allOutFlow = \allToTopkHeight + \allToJoinHeight + \allToLimitHeight + \allToFilterHeight;
\filterInFlow = \allToFilterHeight;
\filterOutFlow = \filterToLimitHeight + \filterToJoinHeight + \filterToTopkHeight;
\limitInFlow = \allToLimitHeight + \filterToLimitHeight;
\limitOutFlow = \limitToJoinHeight;
\joinInFlow = \allToJoinHeight + \limitToJoinHeight + \filterToJoinHeight;
\joinOutFlow = \joinToTopkHeight;
\topkInFlow = \allToTopkHeight + \joinToTopkHeight + \filterToTopkHeight;
}

\def\minBarHeight{7}
\tikzmath{
\allBarHeight = 100;
\filterBarHeight = max(\filterInFlow, \minBarHeight);
\limitBarHeight = max(\limitInFlow, \minBarHeight);
\joinBarHeight = max(\joinInFlow, \minBarHeight);
\topkBarHeight = max(\topkInFlow, \minBarHeight);
}
\tikzmath{
\allBarY = 0;
\filterBarY = -(100 - \filterBarHeight) / 2;
\limitBarY = 24;
\joinBarY = 26;
\topkBarY = (100 - \topkBarHeight) / 2;
}

\def\barWidth{18}
\def\barDistance{50}
\def\boundingBoxYmax{60}
\def\boundingBoxYmin{-60}
\def\boundingBoxXmin{-5}
\def\boundingBoxXmax{225}

    \fill [pattern=north east lines,pattern color=lightgray!70] (46.5,\boundingBoxYmax) rectangle (121.5,\boundingBoxYmin);
    \node[background label] at ({84 + 35}, \boundingBoxYmin+3) {Query\\Compilation};
    \fill [pattern=north west lines,pattern color=lightgray!70] (146.5,\boundingBoxYmax) rectangle (221.5,\boundingBoxYmin);
    \node[background label] at ({184 + 35}, \boundingBoxYmin+3) {Query\\Execution};

    \draw[black,line width=0.5pt] (\boundingBoxXmin,\boundingBoxYmax) -- (\boundingBoxXmax,\boundingBoxYmax) -- (\boundingBoxXmax,\boundingBoxYmin) -- (\boundingBoxXmin,\boundingBoxYmin) -- cycle;
    \node(ylabel)[rotate=90, inner sep=0pt] at (-20, 0) {\small Relative Number of Queries};
    \draw[{Latex}-{Latex},line width=0.5,] ($(ylabel.south)+(4.5pt,-25pt)$) -- ($(ylabel.south)+(4.5pt,25pt)$);
    
    \sankeynode{\allBarHeight}{0}{allqueries}{{0*\barDistance},\allBarY}
    {\tikzset{sankey fill/.append style={snowflakedarkblue}, sankey draw/.append style={snowflakedarkblue}} \sankeyadvance{allqueries}{\barWidth pt} }
    \node[bar label] at (9, \allBarY) {100\%};
    \node[bar title] at (9, \allBarY-\allBarHeight/2) {All};
    \sankeyfork{allqueries}{
        \allToTopkHeight/allqueriestotopkpruningsender,
        \allToJoinHeight/allqueriestojoinpruningsender,
        \allToLimitHeight/allqueriestolimitpruningsender,
        \allToFilterHeight/allqueriestofilterpruningsender}

    \sankeynode{\allToFilterHeight}{0}{filterpruningarrow}{{1*\barDistance-\lineArrowLength}, \filterBarY}
    \sankeyfork{filterpruningarrow}{
        \allToFilterHeight/allqueriestofilterpruningreceiver}
    \sankeynodeend{\allToFilterHeight}{0}{filterpruningarrowcap}{filterpruningarrow}

    \path (allqueriestofilterpruningsender) to[sankey flow] (allqueriestofilterpruningreceiver);
    \node[line label] at (34, -3) {\round{1}{\allToFilter}\%};

    \sankeynode{\filterBarHeight}{0}{filterpruning}{{1*\barDistance}, \filterBarY}
    {\tikzset{sankey fill/.append style={snowflakedarkblue}, sankey draw/.append style={snowflakedarkblue}} \sankeyadvance{filterpruning}{\barWidth pt} }
    \node[bar label] at (59, \filterBarY) {\round{1}{\allFilter}\%};
    \node[bar title] at (59, \filterBarY-\filterBarHeight/2) {Filter};
    \sankeyfork{filterpruning}{
        \filterToLimitHeight/filterpruningtolimitpruningsender,
        \filterToJoinHeight/filterpruningtojoinpruningsender,
        \filterToTopkHeight/filterpruningtotopkpruningsender}

    \sankeynode{\limitInFlow}{0}{limitpruningarrow}{{2*\barDistance-\lineArrowLength}, {{\limitBarY+(\limitBarHeight-\limitInFlow)/2}}}
    \sankeyfork{limitpruningarrow}{
        \allToLimitHeight/allqueriestolimitpruningreceiver,
        \filterToLimitHeight/filterpruningtolimitpruningreceiver}
    \sankeynodeend{\limitInFlow}{0}{limitpruningarrowcap}{limitpruningarrow}

    \sankeyadvance{allqueriestolimitpruningsender}{50pt}
    \path (allqueriestolimitpruningsender) to[sankey flow] (allqueriestolimitpruningreceiver);
    \node[line label,anchor=east] at (85, 32) {\round{4}{\allToLimit}\%};
    \path (filterpruningtolimitpruningsender) to[sankey flow] (filterpruningtolimitpruningreceiver);
    \node[line label,anchor=east] at (85, 21) {\round{1}{\filterToLimit}\%};

    \sankeynode{\limitBarHeight}{0}{limitpruning}{{2*\barDistance}, \limitBarY}
    {\tikzset{sankey fill/.append style={snowflakedarkblue}, sankey draw/.append style={snowflakedarkblue}} \sankeyadvance{limitpruning}{\barWidth pt} }
    \node[bar label] at (109, \limitBarY) {\round{1}{\allLimit}\%};
    \node[bar title] at (109, \limitBarY-\limitBarHeight/2) {LIMIT};
    \sankeyfork{limitpruning}{
        \limitToJoinHeight/limitpruningtojoinpruningsender}
    
    \sankeynode{\joinInFlow}{0}{joinpruningarrow}{{3*\barDistance-\lineArrowLength}, {\joinBarY+(\joinBarHeight-\joinInFlow)/2}}
    \sankeyfork{joinpruningarrow}{
        \allToJoinHeight/allqueriestojoinpruningreceiver,
        \limitToJoinHeight/limitpruningtojoinpruningreceiver,
        \filterToJoinHeight/filterpruningtojoinpruningreceiver}
    \sankeynodeend{\joinInFlow}{0}{joinpruningarrowcap}{joinpruningarrow}

    \sankeyadvance{allqueriestojoinpruningsender}{100pt}
    \path (allqueriestojoinpruningsender) to[sankey flow] (allqueriestojoinpruningreceiver);
    \node[line label] at (112, 46.9) {\round{1}{\allToJoin}\%};
    \sankeyadvance{filterpruningtojoinpruningsender}{50pt}
    \path (filterpruningtojoinpruningsender) to[sankey flow] (filterpruningtojoinpruningreceiver);
    \node[line label] at (112, 4) {\round{1}{\filterToJoin}\%};
    \path (limitpruningtojoinpruningsender) to[sankey flow] (limitpruningtojoinpruningreceiver);
    \node[line label] at (130.5, 30.5) {\round{4}{\limitToJoin}\%};

    \sankeynode{\joinBarHeight}{0}{joinpruning}{{3*\barDistance}, \joinBarY}
    {\tikzset{sankey fill/.append style={snowflakedarkblue}, sankey draw/.append style={snowflakedarkblue}} \sankeyadvance{joinpruning}{\barWidth pt} }
    \node[bar label] at (159, \joinBarY) {\round{1}{\allJoin}\%};
    \node[bar title] at (159, \joinBarY-\joinBarHeight/2) {Join};
    \sankeyfork{joinpruning}{
        \joinToTopkHeight/joinpruningtotopkpruningsender}
             
    \sankeynode{\topkInFlow}{0}{topkpruningarrow}{{4*\barDistance-\lineArrowLength}, {\topkBarY+(\topkBarHeight-\topkInFlow)/2}}
    \sankeyfork{topkpruningarrow}{
        \allToTopkHeight/allqueriestotopkpruningreceiver,
        \joinToTopkHeight/joinpruningtotopkpruningreceiver,
        \filterToTopkHeight/filterpruningtotopkpruningreceiver}
    \sankeynodeend{\topkInFlow}{0}{topkpruningarrowcap}{topkpruningarrow}

    \path (allqueriestotopkpruningsender) to[sankey flow] (allqueriestotopkpruningreceiver);
    \node[line label] at (162, 54) {\round{2}{\allToTopk}\%};
    \sankeyadvance{filterpruningtotopkpruningsender}{100pt}
    \path (filterpruningtotopkpruningsender) to[sankey flow] (filterpruningtotopkpruningreceiver);
    \node[line label] at (162, -5) {\round{2}{\filterToTopk}\%};
    \path (joinpruningtotopkpruningsender) to[sankey flow] (joinpruningtotopkpruningreceiver);
    \node[line label] at (173, 41) {\round{3}{\joinToTopk}\%};

    \sankeynode{\topkBarHeight}{0}{topkpruning}{{4*\barDistance}, \topkBarY}
    {\tikzset{sankey fill/.append style={snowflakedarkblue}, sankey draw/.append style={snowflakedarkblue}} \sankeyadvance{topkpruning}{\barWidth pt} }
    \node[bar label] at (209, \topkBarY) {\round{1}{\allTopk}\%};
    \node[bar title] at (209, \topkBarY-\topkBarHeight/2) {Top-k};

  \end{sankeydiagram}
\end{tikzpicture}

    \caption{Flow diagram visualizing how many queries are subject to which pruning technique(s). Pruning techniques are executed from left to right in order. Queries are based on a representative workload sample across all customers from Mar 25\textsuperscript{th} to Mar 28\textsuperscript{th} 2025.}
    \label{fig:pruning-flow}
\end{figure}
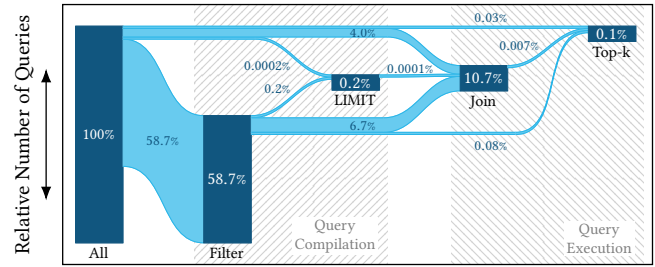

\section{Discussion}
\label{s:discussion}

In the following sections, we will discuss how pruning can be done for open table formats like Apache Iceberg, extend and discuss the idea of \emph{predicate caching}, and lastly show that synthetic benchmarks do not accurately capture the importance of pruning.

\subsection{Pruning for Iceberg Tables in a Data Lake}
\label{s:pruning-for-iceberg}

Besides its highly optimized proprietary format, Snowflake also supports the Apache Iceberg~\cite{icebergformat} open table format backed by the Apache Parquet file format~\cite{parquetformat}.
Snowflake’s query engine seamlessly handles both formats, with pruning techniques operating transparently across them.
Consequently, all pruning methods discussed in this paper apply equally to Iceberg tables in Snowflake.
Similar to Snowflake's own file format, the Apache Parquet file format also follows a PAX-style layout~\cite{ailamaki2001weaving}, allowing columnar metadata that can be used for pruning on row-group level.
Parquet files may also include page-level indexes~\cite{parquetpageindex} and manifest files in Apache Iceberg can further contain metadata on file-level, which is also utilized by Snowflake for pruning.

Metadata is the cornerstone of pruning---without it, no pruning is possible and therefore the quality and correctness of metadata plays a key role in query performance.
Consequently, backfilling missing metadata is crucial for analytical systems operating on data lakes.
If a Parquet file contains metadata, Snowflake can immediately use it for pruning.
However, if there is no metadata, Snowflake can reconstruct it by performing a full table scan to compute missing metadata entries, which can then be used for subsequent queries.
Similarly, if an Iceberg table does not contain metadata in manifest files, Snowflake can reconstruct it using the metadata from the underlying Parquet files.

\subsection{What about Predicate Caching?}\label{s:topk-against-predicate-caching}
\begin{figure}[t]
    \centering
    \begin{subfigure}[t]{0.46\linewidth}
        \includegraphics[width=\linewidth]{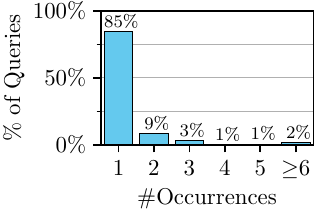}
        \caption{Over 3 days}
        \label{fig:sub:topk-repetitiveness-3days}
    \end{subfigure}
    \hfill
    \begin{subfigure}[t]{0.46\linewidth}
        \includegraphics[width=\linewidth]{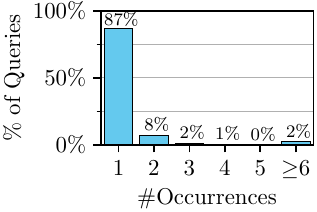}
        \caption{Over 1 month}
        \label{fig:sub:topk-repetitiveness-1month}
    \end{subfigure}
    \caption{Repetitiveness of top-k query plan shapes in Snowflake over a 3-day and 1-month period. Most query plan shapes appear only once. Queries are based on a representative workload across all customers from Nov 5\textsuperscript{th} to Nov 7\textsuperscript{th} 2024, and from Oct 8\textsuperscript{th} to Nov 7\textsuperscript{th} 2024, respectively.}
    \label{fig:topk-repetitiveness}
\end{figure}

Schmidt et al. recently proposed \emph{predicate ca\-ching}~\cite{Schmidt2024}.
While they focus on caching relevant partitions for filters only, an extension to top-k queries can be envisioned as follows:

For repetitive top-k queries (see \Cref{fig:topk-repetitiveness}), the micro-partitions contributing to the final top-k result can be identified by recording partition information alongside each tuple in the top-k heap during query processing.
This list of contributing micro-partitions can then be stored in a global "predicate cache."
When the same top-k query is executed again, the query engine performs a cache lookup, and if a matching entry exists, it processes only the cached micro-partitions.
If predicate caching was perfect---yielding no false-positive partitions---the query engine would scan only the partitions necessary to produce the top-k result.
This approach could outperform our pruning-based method, especially for randomly sorted datasets with mostly overlapping min/max ranges where pruning may struggle to exclude many partitions.

\noindent However, due to the space limitations of the predicate cache, the caching result might include a significant number of false-positive partitions, especially for large datasets.
For large, (partially) sorted tables, our pruning-based method can more effectively exclude non-relevant micro-partitions, likely outperforming predicate caching.

To maximize efficiency, we argue that both techniques should be implemented, as their respective strengths can complement each other and address their individual shortcomings.

The integration of predicate caching with top-k queries introduces additional complexities when dealing with \texttt{UPDATE}s and \texttt{DELETE}s.
While Schmidt\etal's predicate caching for filters supports \texttt{INSERT}, \texttt{UPDATE}, and \texttt{DELETE} operations on the base table, the same flexibility does not fully extend to top-k queries.

If a row in the top-k result is deleted, another row must take its place. 
However, there is no guarantee that this replacement row (the
$k+1$-th row) resides within the cached micro-partitions, potentially leading to incorrect results.
Similarly, \texttt{UPDATE}s to the ordering column may reorder rows in a way that invalidates the cached partitions.
In contrast, \texttt{UPDATE}s to non-ordering columns and \texttt{INSERT}s are safe and do not affect the correctness of the cache.

In contrast, our pruning-based approach for top-k queries is naturally robust against DML operations that modify the ordering column, maintaining correctness without the need for cache invalidation, making it a robust choice particularly for dynamic datasets.
Additionally, we showed that top-k queries are generally not particularly repetitive  (see \Cref{fig:topk-repetitiveness}), which diminishes the potential of \textit{predicate caching}.
In contrast to that, our approach also supports ad-hoc top-k queries.

\subsection{Drawing a Line to TPC-H}
\label{s:tpch}

\begin{figure}[tb]
    \centering
    \includegraphics[width=\linewidth]{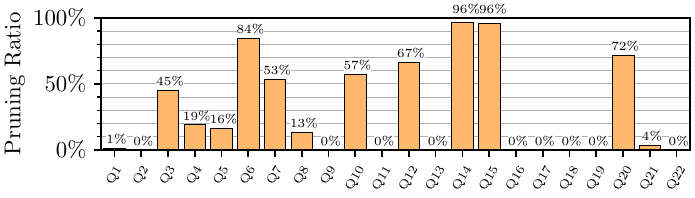}
    \caption{Pruning ratios of queries in TPC-H SF100, clustered on \texttt{l\_shipdate} and \texttt{o\_orderdate}, executed on an XSMALL virtual warehouse in Snowflake as of March 2025. Average pruning ratio over the whole workload is 28.7\%, with a median per-query pruning ratio of 8.3\%.}
    \label{fig:tpch-pruning-ratio}
\end{figure}

Comparing the pruning capabilities of real-world customer workloads to artificial benchmarks like TPC-H showed some stark differences that are briefly discussed in the following.
We ran TPC-H SF100 on an XSMALL virtual warehouse in Snowflake to ensure pruning behavior was not influenced by effects of data placement across different machines.
We clustered the dataset by \texttt{l\_shipdate} and \texttt{o\_orderdate} to allow better partition pruning and exploitation of column correlations as discussed by Dre\-se\-ler\etal~\cite{dreseler2020quantifying}.
\Cref{fig:tpch-pruning-ratio} shows the pruning ratios of TPC-H SF100 queries in this setup.
Notably, these ratios are significantly lower than those observed for queries executed on Snowflake.
As most pruning came from filter pruning on tables \texttt{LINEITEM} and \texttt{ORDERS} and no pruning happened with default data clustering, it is clear that this
discrepancy comes from the relatively low selectivity of query predicates in TPC-H compared to real-life workloads.
Further, no query in TPC-H was able to exploit top-k pruning and due to the deterministic nature of the benchmark, no \texttt{LIMIT} pruning was possible. Join pruning was also greatly underrepresented.

The key takeaway is that the impact of partition pruning is not adequately captured by standard synthetic benchmarks like TPC-H, reinforcing our earlier claim about the challenges of accurately evaluating pruning techniques.
We are not the first to notice this. Previous work also aimed to address similar issues with synthetic benchmarks~\cite{ding2021dsb, wan2023stitcher}.
While synthetic benchmarks can approximate real-world pruning behavior, careful query design is essential.
We identify the key aspect to be a more realistic selectivity of predicates. Further, enough opportunity for join pruning needs to exist, \eg by designing very small build sides or having a sufficient correlation in data layout between build and probe sides. Lastly, such a benchmark should also contain non-deterministic \texttt{LIMIT} queries.

\section{Conclusion}
\label{s:conclusion}

In this paper, we presented a comprehensive analysis of partition pruning techniques of the Snowflake Data Platform.
We explained the functioning and evaluated the impact and interplay of four pruning strategies: filter pruning, \texttt{LIMIT} pruning, top-k pruning, and \texttt{JOIN} pruning.
Our findings, based on real-world customer workloads, demonstrate the substantial performance benefits these techniques offer, in particular for cloud-based systems.
Every presented pruning technique achieves substantial pruning ratios for applicable queries with 99\% for filter pruning, 70\% for \texttt{LIMIT} pruning, 77\% for top-k pruning, and 79\% for join pruning.
Overall, pruning is one of the enablers of cloud-based data processing with Snowflake pruning 99.4\% of micro-partitions across all queries.

We further showed that synthetic benchmarks do not accurately capture the importance of partition pruning for query performance.
By sharing our experiences with Snowflake's pruning mechanisms, we aim to inspire further research and innovation in this area.

\begin{acks}
We gratefully acknowledge the contributions of the many members of the Snowflake team, whose collective efforts formed the foundation of this paper. We specifically wish to thank Abdul Q Munir, Hossein Ahmadi, Berni Schiefer, Thierry Cruanes, Chris Baynes, and Changbin Song for their invaluable support of this work.
\end{acks}

\bibliographystyle{ACM-Reference-Format}
\newcommand{\shownote}[1]{\unskip}
\bibliography{literature}

\end{document}